\documentclass[journal=jctcce,manuscript=article]{achemso}
\setkeys{acs}{usetitle = true} % introduce title in the references
\usepackage[version=3]{mhchem} % Formula subscripts using \ce{}
\usepackage[T1]{fontenc}       % Use modern font encodings
\usepackage{amsmath,amssymb,latexsym,epsfig,multirow,tabularx}
\usepackage{amsthm}
\usepackage{verbatim}
\usepackage{siunitx} % unit package , usage: $\SI{0.5}{\angstrom}$
\usepackage{booktabs} %nice table
\usepackage{longtable} % longtable
\usepackage{graphicx}
\usepackage[tableposition=t]{caption} % keep caption at normal size
\usepackage{subcaption}
\usepackage[titletoc,toc,title]{appendix}
\usepackage{threeparttable} %add footnote underneath the table
\usepackage{wrapfig} % to wrap text around figure
\usepackage{color}
\usepackage[normalem]{ulem} %strikeout a word
\author{Duc D. Nguyen}\altaffiliation{The first two authors contribute equally to this work}
\author{Bao Wang}\altaffiliation{The first two authors contribute equally to this work}
\author{Guo-Wei Wei}
\email{wei@math.msu.edu}
\affiliation
{Department of Mathematics\\
Michigan State University, MI 48824, USA}
\alsoaffiliation
{Department of Electrical and Computer Engineering\\
Michigan State University, MI 48824, USA}
\alsoaffiliation
{Department of Biochemistry and Molecular Biology\\
Michigan State University, MI 48824, USA}

\title{Accurate, robust and reliable calculations of Poisson-Boltzmann
 binding energies}


\begin{document}

%\begin{tocentry}
%
%Some journals require a graphical entry for the Table of Contents.
%This should be laid out ``print ready'' so that the sizing of the
%Inside the \texttt{tocentry} environment, the font used is Helvetica
%8\,pt, as required by \emph{Journal of the American Chemical
%The surrounding frame is 9\,cm by 3.5\,cm, which is the maximum
%permitted for  \emph{Journal of the American Chemical Society}
%graphical table of content entries. The box will not resize if the
%content is too big: instead it will overflow the edge of the box.
%
%This box and the associated title will always be printed on a
%separate page at the end of the document.
%
%\end{tocentry}

\begin{abstract}
%{\centering
%    	\includegraphics[width=0.38\columnwidth]{1b27_fixed.pdf}
%    	\includegraphics[width=0.32\columnwidth]{bd_total_12.pdf}
%\par
%}
Poisson-Boltzmann (PB) model is   one of the most popular implicit solvent models in biophysical modeling and computation. The ability of providing accurate and reliable PB estimation of  electrostatic solvation free energy, $\Delta G_{\text{el}}$, and binding free energy, $\Delta\Delta G_{\text{el}}$,   is      important to computational biophysics and biochemistry.  Recently, it has been   warned in the literature (Journal of Chemical Theory and
Computation 2013, 9, 3677-3685) that the widely used grid spacing of $\SI{0.5}{\angstrom}$ produces unacceptable errors in  $\Delta\Delta G_{\text{el}}$ estimation  with the solvent exclude surface (SES). In this work, we investigate the   grid dependence of our  PB solver (MIBPB) with SESs for estimating both electrostatic solvation  free energies and electrostatic  binding free energies. It is found that the relative absolute error of  $\Delta G_{\text{el}}$ obtained at the grid spacing of $\SI{1.0}{\angstrom}$ compared to $\Delta G_{\text{el}}$ at  $\SI{0.2}{\angstrom}$ averaged  over 153 molecules is less than 0.2\%. Our results indicate that  the use of grid spacing $\SI{0.6}{\angstrom}$ ensures accuracy and reliability in  $\Delta\Delta G_{\text{el}}$ calculation. In fact, the grid spacing of $\SI{1.1}{\angstrom}$ appears to deliver  adequate accuracy for high throughput screening.

\end{abstract}
\maketitle

%\newpage
\section{Introduction}
 Electrostatics is ubiquitous in biomolecular and cellular systems and of paramount importance to biological processes. Accurate and reliable  prediction of electrostatic binding free energy, $\Delta\Delta G_{\text{el}}$, is crucial to biophysical modeling and computation. The prediction of  $\Delta\Delta G_{\text{el}}$  plays a vital role in the study of many cellular processes, such as signal transduction, gene expression, and protein synthesis. Additionally, many pharmaceutical applications, specially in the final stage of the drug design, rely on the accurate and reliable calculation of binding free energy. Technically, the accuracy and reliability of electrostatic binding energy prediction depend essentially on the quality of electrostatic solvation ($\Delta G_{\text{el}}$)  estimation, which can be achieved by solving the Poisson-Boltzmann (PB) equation in the implicit solvent model \cite{Gilson:1987,Gilson:1993,Talley:2008, Zhou:2008b,RenP:2012}. In past decades, the development of a robust PB solver catches much attention in computational biophysics and biochemistry. Mathematically, most PB solvers reported in the literature are based on three major approaches, namely, the finite difference method (FDM) \cite{Jo:2008}, the finite element method (FEM) \cite{Baker:2001b}, and the boundary element method (BEM) \cite{Geng2013:3,LuBenzhuo:2013}. Among them, the FDM is prevalently used in the field due to its simplicity in implementation. The emblematic solvers in this category are Amber PBSA \cite{PBSA:2008,PBSA:2009}, Delphi \cite{Delphi:2012,Rocchia:2002}, APBS \cite{Baker:2001} and CHARMM PBEQ \cite{Jo:2008}.

Recently, it    has been  warned that ``the widely used grid spacing of 0.5 \AA~ produces unacceptable errors in  $\Delta\Delta G_{\text{el}}$''  \cite{Harris:2013}.
 Although all results were obtained with the  adaptive Cartesian grid (ACG) finite difference PB equation solver  \cite{Boschitsch:2015} in this work, similar results were reported in a later 
 study  \cite{Sorensen:2015} by using APBS, DelPhi and PBSA. 
Therefore, these studies have arisen serious concerns about the validity of using PB model for biomolecular electrostatic binding analysis at an affordable grid spacing of 0.5 \AA.   

In the past few years,  there have been many attempts to develop highly accurate PB solvers using advance techniques for interface treatments \cite{Boschitsch:2015,Fenley:2015b,CWang:2013,Mirzadeh:2011}.  The later verison of the ACG solver   \cite{Boschitsch:2015,Fenley:2015b} has somewhat remedied the grid-dependence issue for estimates of binding energy. However, no confirmation for the reliable use of grid spacing of 0.5 \AA~ in  $\Delta\Delta G_{\text{el}}$ has been given. 
In this work, we investigate the grid dependence of our PB solver (MIBPB) \cite{DuanChen:2011a,BaoWang2015PB}    in estimating both electrostatic solvation free energies and electrostatic binding free energies. The MIBPB solver is by far the only existing method that is second-order accurate in $L_{\infty}$ norm for solving the Poisson-Boltzmann equation with discontinuous dielectric constants, singular charge sources, and geometric singularities from the solvent excluded surfaces (SESs) of biomolecules \cite{DuanChen:2011a}. Here the $L_{\infty}$ norm means the maxmum absolute error measure   and   ``second order accurate'' means that the error reduces four times when the grid spacing  is halved.   Contrary to the findings in the literature  \cite{Harris:2013},    our results indicate that the use of grid spacing 0.6 \AA~ ensures accuracy and reliability in $\Delta\Delta G_{\text{el}}$ calculation. In fact, a grid spacing of 1.1 \AA~ appears to deliver adequate accuracy for   high throughput screening. We therefore believe that when it is used properly, the PB methodology is able to deliver accurate and reliable electrostatic binding analysis.

\section{Methods}

\subsection{MIBPB   package}

In the current work, we employ the our MIBPB package \cite{DuanChen:2011a,BaoWang2015PB}   to predict the electrostatic solvation free energy. The MIBPB  package is  a second-order convergence PB solver for dealing with the  SESs of biomolecules. Numerically, there are three major obstacles in constructing accurate and reliable PB solvers. First, commonly used solvent-solute interfaces, i.e., the van der Walls (vdW) surface, solvent accessible surface (SAS), and the solvent excluded surface (SES) \cite{Richards:1977,Connolly:1983} admit geometric singularities, such as sharp tips, cusps and self-intersecting surfaces \cite{Sanner:1996}, which make the rigorous enforcement of interface jump conditions a formidable task in PB solvers. An advanced mathematical interface techniques, the matched interface and boundary (MIB) method \cite{Zhou:2006d,Zhou:2006c,Yu:2007,Yu:2007c,Yu:2007a,KLXia:2014e}, is employed in the MIBPB package to achieve the second order accuracy in handling biomolecular SESs.  Additionally, the atomic singular charges described by the delta functions give rise to another difficulty in constructing highly accurate PB solver. A  Dirichlet-to-Neumann map technique has been developed in the MIBPB package to avoid the numerical approximation of singular charges by using the analytical Green's functions \cite{Geng:2007a}. Finally, the nonlinear   Boltzmann term can affect solver efficiency when handled inappropriately, particularly for BEMs. A quasi-Newton algorithm is implemented in the   MIBPB package \cite{DuanChen:2011a,BaoWang2015PB} to take care the nonlinear term  \cite{DuanChen:2011a,BaoWang2015PB}.

\subsection{Interface generation}

Many studies suggest that SES is able to deliver the state of the art accurate modeling of the solvated molecule  \cite{Baker:2001b,Rocchia:2002,LuBenzhuo:2013}. As a result, much effort has been paid to developing an accurate and robust SES software \cite{Sanner:1996,Decherchi:2013}. However, the MSMS software  \cite{Sanner:1996} generates a Lagrangian  representation of the SES and is inconvenient for the Cartesian domain implementation of PB solvers. A Lagrangian to Eulerian transformation is required to convert MSMS surfaces for our Cartesian based MIBPB solver \cite{Zhou:2008b}.  Most recently, we have developed a new SES software,  Eulerian solvent excluded surface (ESES), to directly generate the SESs in the Eulerian representation  \cite{ESES:2015}. Our ESES software enables the MIBPB solver to produce a reliable $\Delta G_{\text{el}}$. Both MSMS and ESES are supported by our MIBPB software. By increasing the MSMS surface density, the electrostatic solvation free  energies calculated by using MSMS converge to those obtained by using ESES \cite{ESES:2015}. Therefore, only results employing ESES are shown in this work.

\subsection{Data sets}
In the present work, we adopt three sets of biomolecular complexes employed in the literature \cite{Harris:2013} for  solvation free energy and binding free energy estimations. Specifically, the first set, Data Set 1, is a collection of DNA-minor groove drug complexes having a narrow range of $\Delta\Delta G$. The Protein Data Bank (PDB) IDs (PDBIDs) for this set are as follows: 102d, 109d, 121d, 127d, 129d, 166d, 195d, 1d30, 1d63, 1d64, 1d86, 1dne, 1eel, 1fmq, 1fms, 1jtl, 1lex, 1prp, 227d, 261d, 164d, 289d, 298d, 2dbe, 302d, 311d, 328d, and 360d. The second set, Data Set 2, includes various wild-type and mutant barnase-barstar complexes. Its PDBIDs are as follows: 1b27, 1b2s, 1b2u, 1b3s, 2aza4, 1x1w, 1x1y, 1x1u, and 1x1x. In the last set, Data Set 3, we investigate RNA-peptide complexes with following PDBIDs: 1a1t, 1a4t, 1biv, 1exy, 1g70, 1hji, 1i9f, 1mnb, 1nyb, 1qfq, 1ull, 1zbn, 2a9x, and 484d. The detail of the structural prepossessing  can be found in Ref. \citeauthor{Harris:2013}. 
The data sets can be downloaded from website {http://www.sb.fsu.edu/\~{}mfenley/convergence/downloads/convergence\_pqr\_sets.tar.gz.} They are also available from the present authors upon request.  

%Figure  \ref{fig.1b27}, generated by Chimera software \cite{UCSFChimera:2004}, describes the structural response to mutation at a protein-protein interface with PDBID 1b27.

%\begin{figure}[!tb]
	%\begin{center}
		%\includegraphics[width=0.50\columnwidth]{1b27_fixed.pdf}
		%\caption{Structural response to mutation at a protein-protein interface (PDBID: 1b27).}
		%\label{fig.1b27}
	%\end{center}
%\end{figure}

\subsection{Poisson-Boltzmann calculation details}

The electrostatics binding free energy is a measure of binding affinity of two compounds due to the electrostatics interaction. Based on the free energy cycle, the electrostatics binding free energy can be calculated by the following formula \cite{Marcia:2011BiophysicalChemistry}
\begin{equation}
\Delta\Delta G_{\text{el}}=\left(\Delta G_{\text{el}}\right)_{\text{AB}} - \left(\Delta G_{\text{el}}\right)_{\text{A}} - \left(\Delta G_{\text{el}}\right)_{\text{B}} + \left(\Delta\Delta G_{\text{el}}\right)_{\text{Coulomb}} ,
\end{equation}
where $(\Delta G_{\text{el}})_{\text{AB}}$ is the electrostatic solvation free energy of the bounded complex AB, $(\Delta G_{\text{el}})_{\text{A}}$ and $(\Delta G_{\text{el}})_{\text{B}}$ are the electrostatic solvation free energies of the unbounded components A and B, and $(\Delta\Delta G_{\text{el}})_{\text{Coulomb}}$ is the electrostatic binding free energy of the two components in vacuum.

The electrostatic solvation free energies $\Delta G_{\text{el}}$  are obtained by using MIBPB software \cite{DuanChen:2011a,BaoWang2015PB} while the binding energy  $(\Delta\Delta G_{\text{el}})_{\text{Coulomb}}$ is easily evaluated analytically via the following formula
\begin{equation}
 \left(\Delta\Delta G_{\text{el}}\right)_{\text{Coulomb}}=\sum_{i,j}\frac{q_i q_j}{\epsilon_m r_{ij}},\quad
 \forall i\in \text{A}, j\in\text{B},
\end{equation}
where $q_i$ and $q_j$ are the corresponding charges of the given pair of atoms, and $r_{ij}$ is the distance between this pair. Here, $\epsilon_m$ is the dielectric constant of the solute region. Table S3 (in the supporting information) lists $(\Delta\Delta G_{\text{el}})_{\text{Coulomb}}$ values of 51 studied complexes.

In all our calculations, the absolute temperature of the ionic solvent is chosen to be $T=298\text{~K}$, the dielectric constants for solute and solvent are 1 and 80, and the ionic strength is $0.1$ M NaCl. The PBE is solved by the linearized solver, but the nonlinear one does not produce any notably differences. The incomplete LU biconjugate gradient squared (ILUBGS) solver is employed to solve all linear systems risen by the MIBPB approach. To maintain consistent computations of the PB solver at different grid sizes, the criteria convergence of ILUBGS solver measured by $L_2$-norm is set to be $10^{-6}$, and the maximum iteration number is set to 100,000. The predictions of MIBPB solver on $\Delta G_{\text{el}}$ and $\Delta\Delta G_{\text{el}}$ are confirmed  by other solvers such as PBSA \cite{PBSA:2008,PBSA:2009}, Delphi \cite{Delphi:2012,Rocchia:2002}, and APBS \cite{Baker:2001} at the grid size of 0.2 \AA, see Table S2 of Supporting Information.

\begin{figure}[!tb]
	\centering
	\begin{subfigure}[t]{0.20\textwidth}
		\centering
		\includegraphics[width=1.1\textwidth]{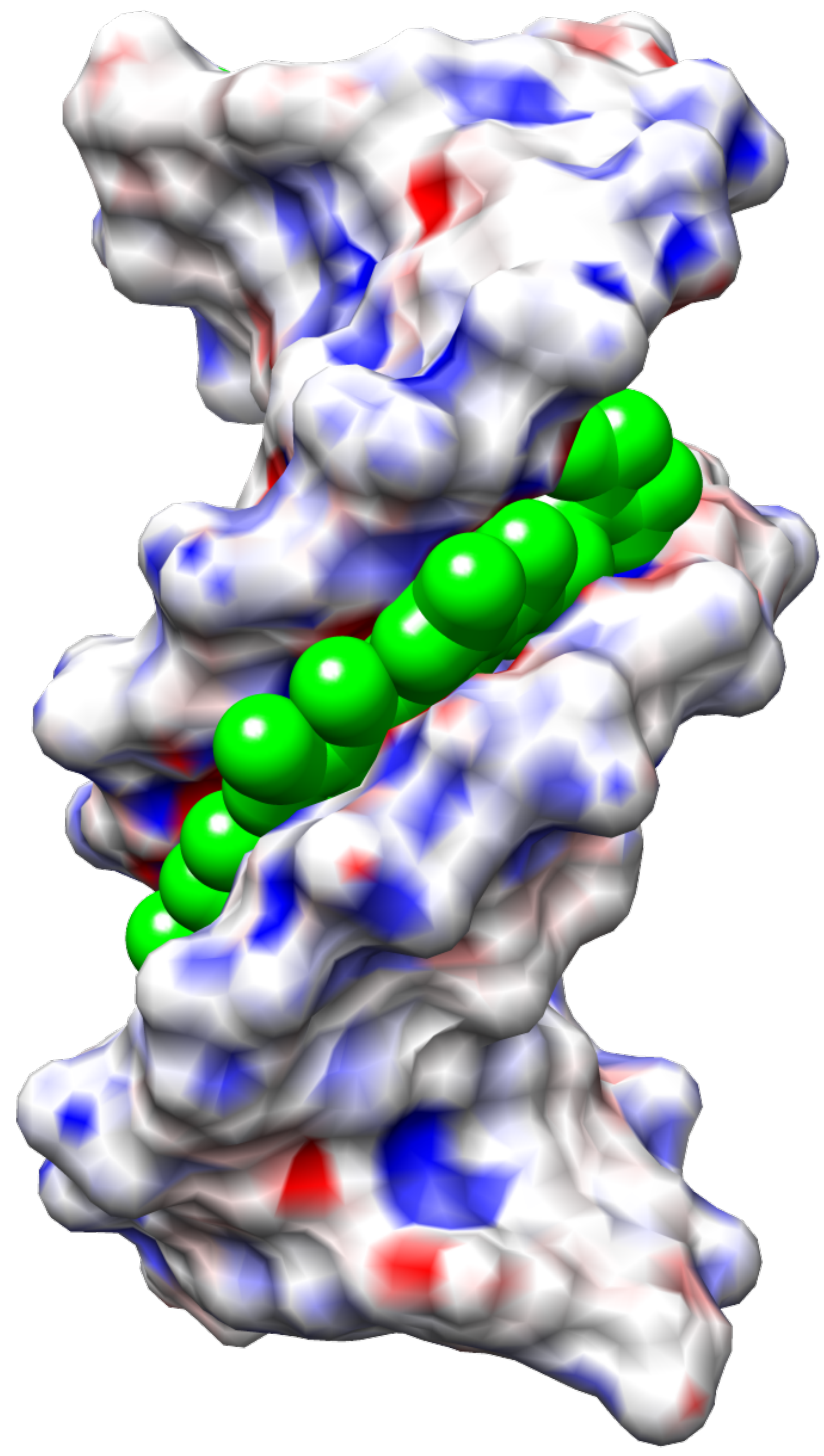}
		\caption{PDBID: 121d}
	\end{subfigure}
\hspace{10pt}
~
	\begin{subfigure}[t]{0.31\textwidth}
		\centering
		\vspace*{-30ex}
	\includegraphics[width=1.1\textwidth]{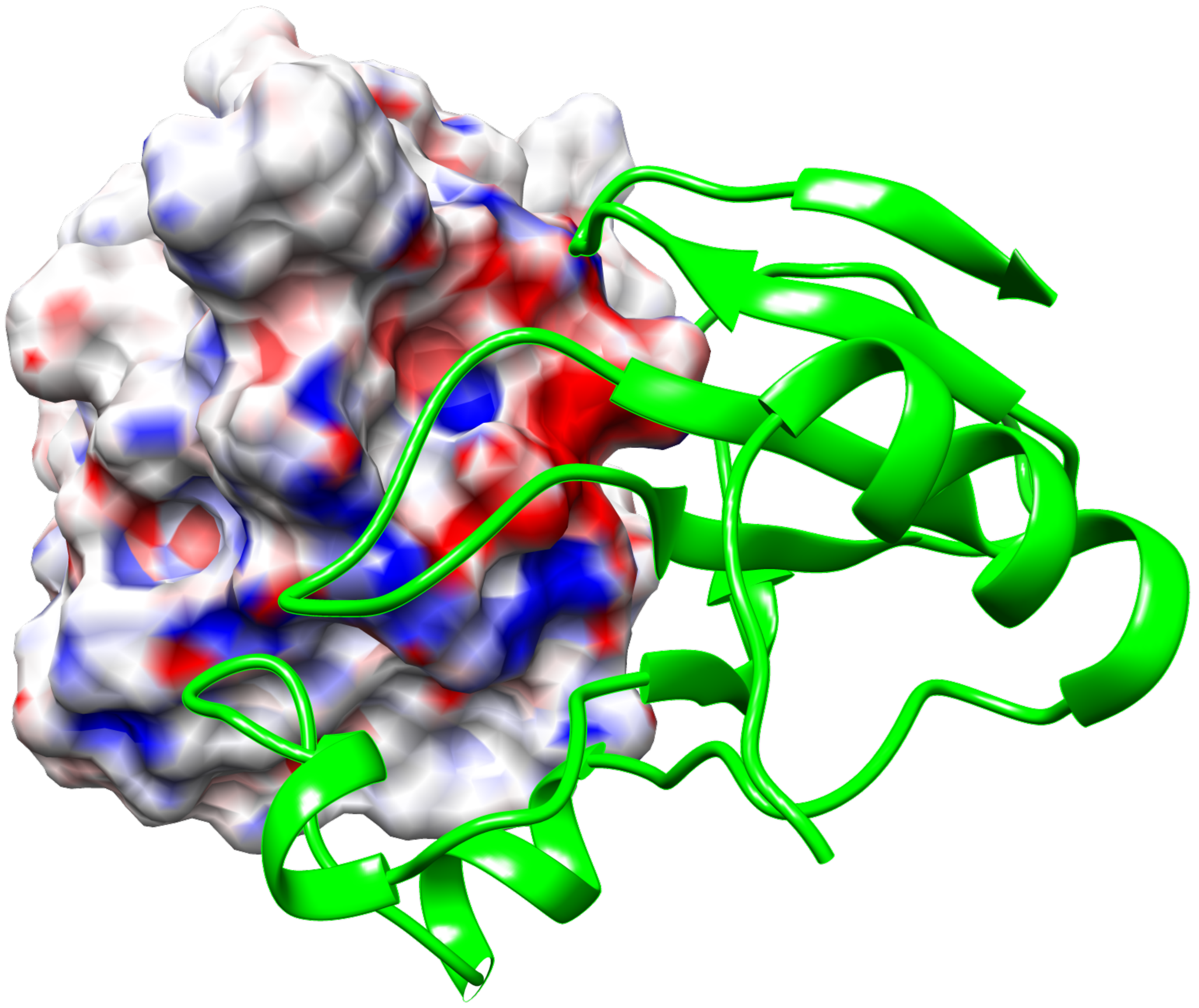}
		\vspace*{1ex}
				\caption{PDBID: 1b3s}
	\end{subfigure}	
	\hspace{2pt}
~~~
\begin{subfigure}[t]{0.35\textwidth}
		\centering
		\vspace*{-35ex}
		\includegraphics[width=1.0\textwidth]{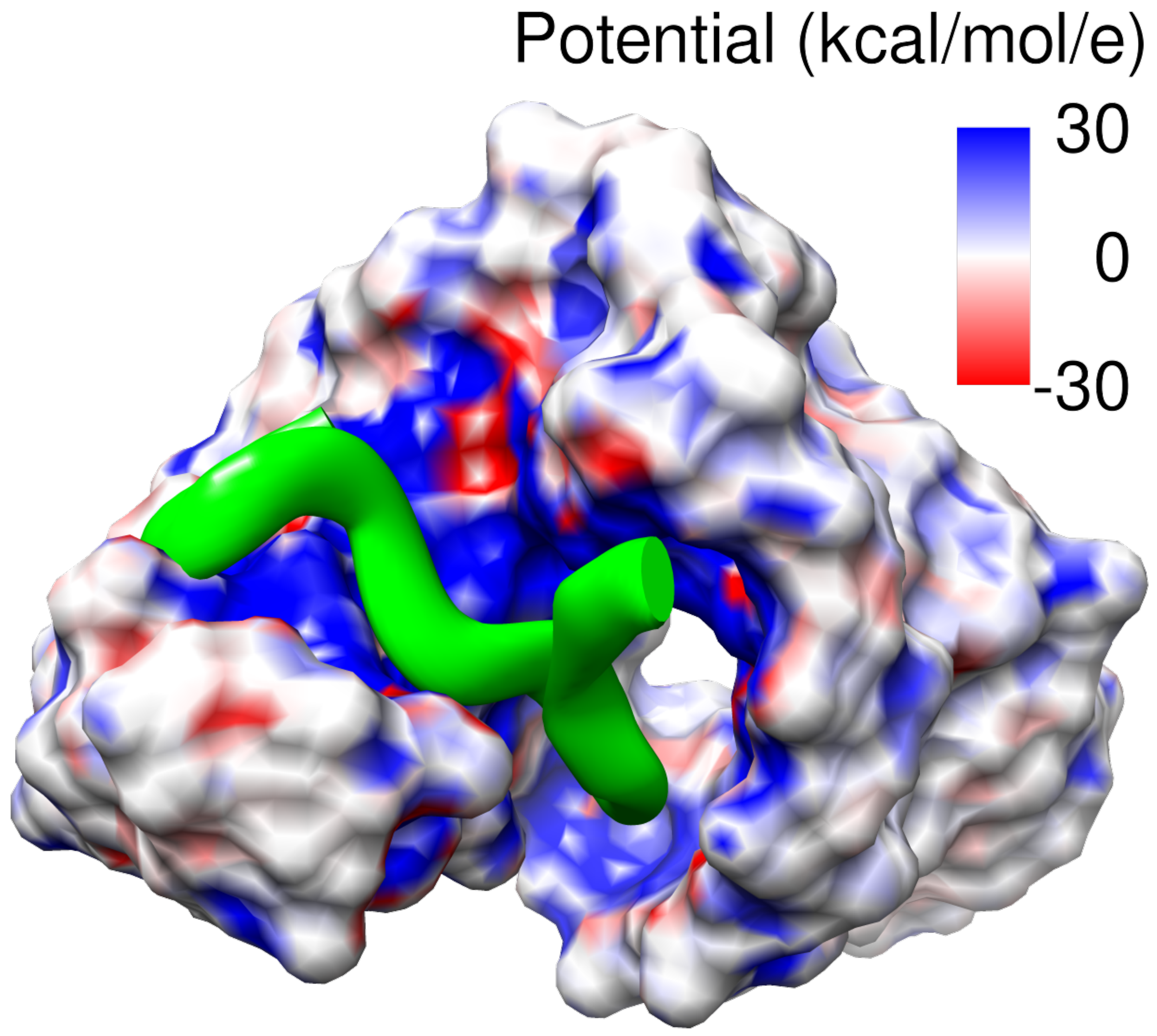}
		\vspace*{3ex}
		\caption{PDBID: 1biv}
	\end{subfigure}
	\caption{Illustration of surface electrostatic potentials (in units of kcal/mol/e) for three complexes, generated by Chimera software \cite{UCSFChimera:2004}. (a) PDBID: 121d (in Drug-DNA complexes); (b) PDBID: 1b3s (in barnase-barstar complexes); (c) PDBID: 1biv (in RNA-peptide complexes).   }
		\label{fig.ele_pot}
\end{figure} 

\section{Results and discussion}

As described above, we consider three sets of binding complexes, namely,  drug-DNA, barnase-barstar and RNA-peptide systems.  For the sake of illustration,  three sample surface electrostatic potentials, each from one distinct set, are depicted in   Fig. \ref{fig.ele_pot}. PDBIDs for these three complexes are respectively 121d (in Drug-DNA complexes), 1b3s (in barnase-barstar complexes), and 1biv (in RNA-peptide complexes). In the rest of this section, we explore the  
influence of grid spacing in Poisson-Boltzmann equation solvation and binding free energy estimations using our MIBPB solver.

\subsection{The influence of grid spacing in $\Delta G_{\text{el}}$ estimation}

\begin{figure}[!tb]
	\begin{center}
		\includegraphics[width=0.30\columnwidth]{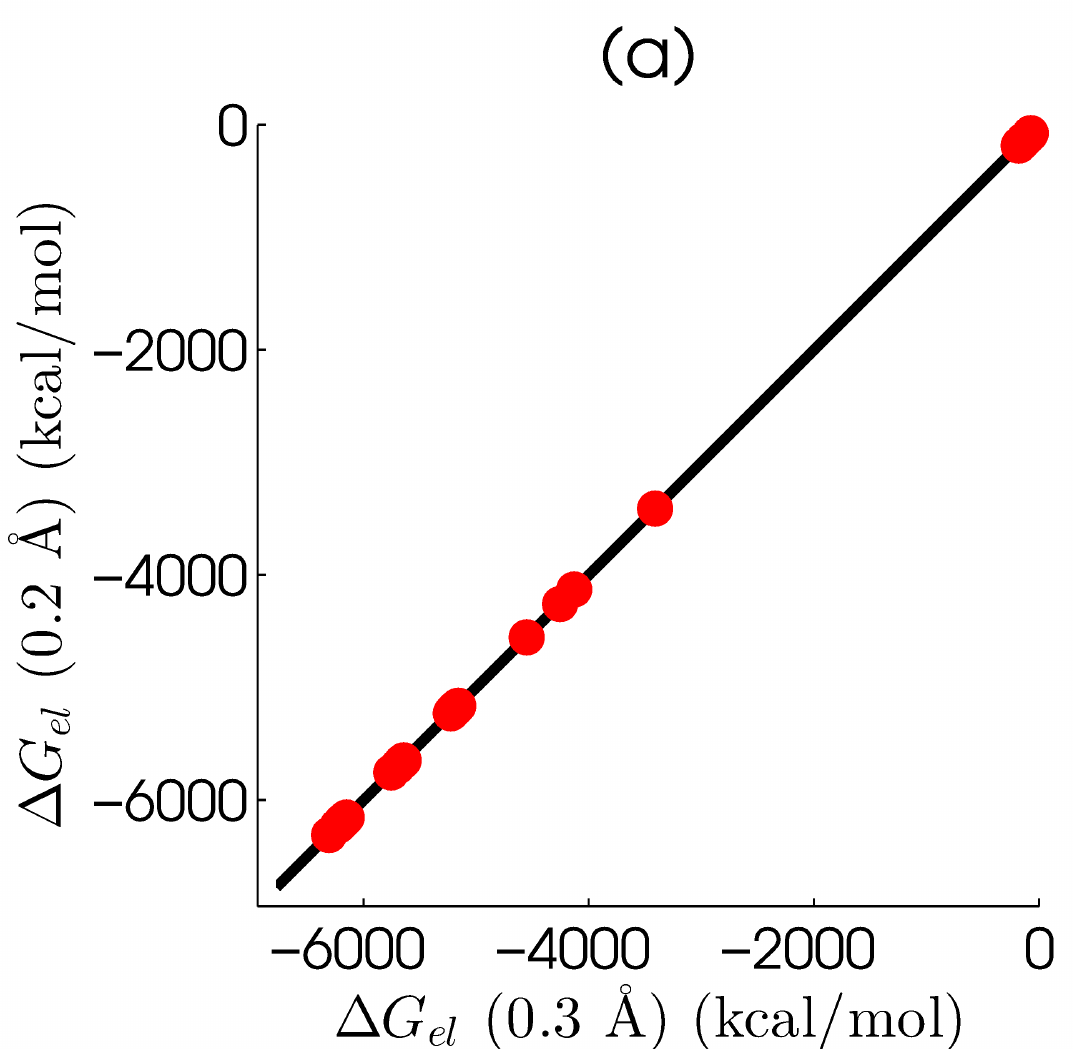}\quad
		\includegraphics[width=0.30\columnwidth]{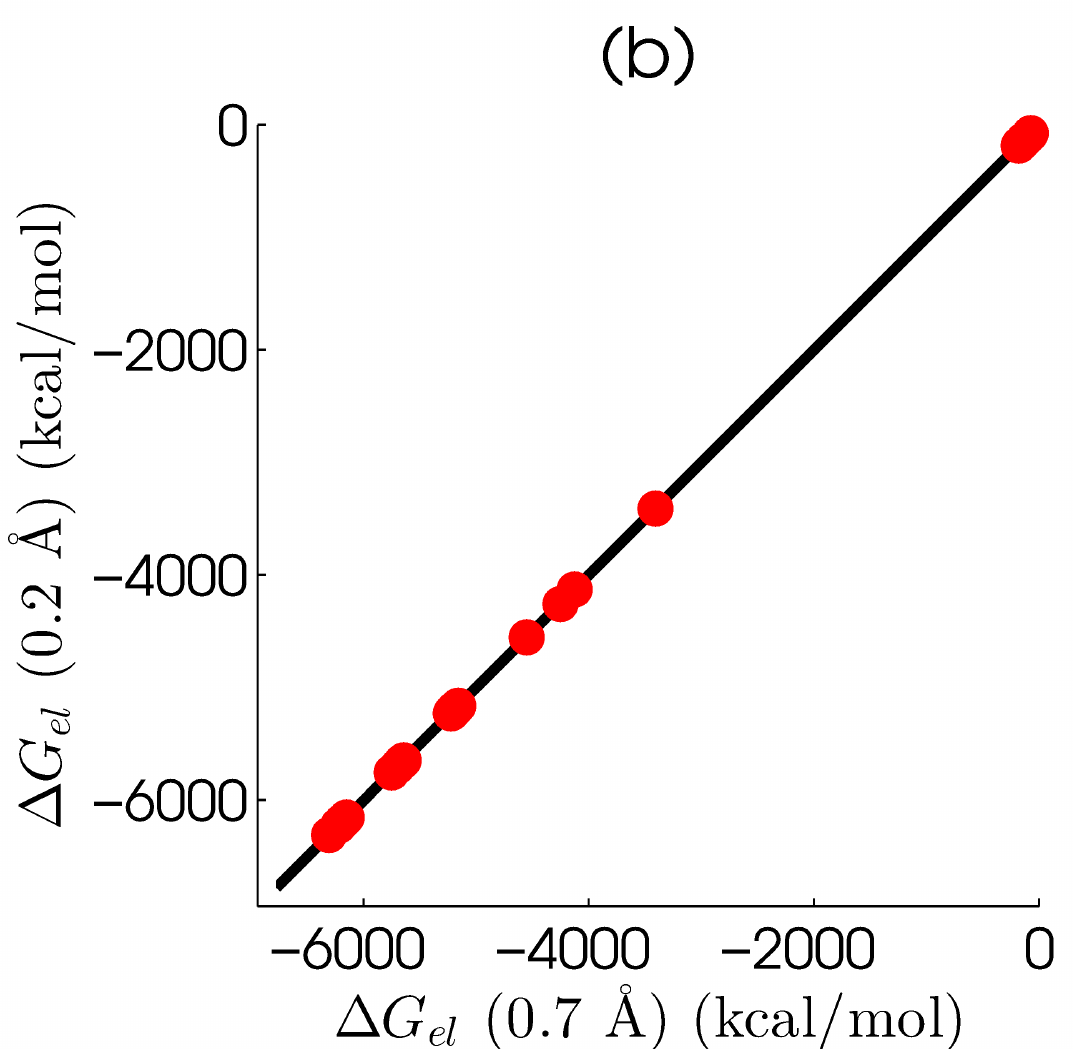}\quad
		\includegraphics[width=0.30\columnwidth]{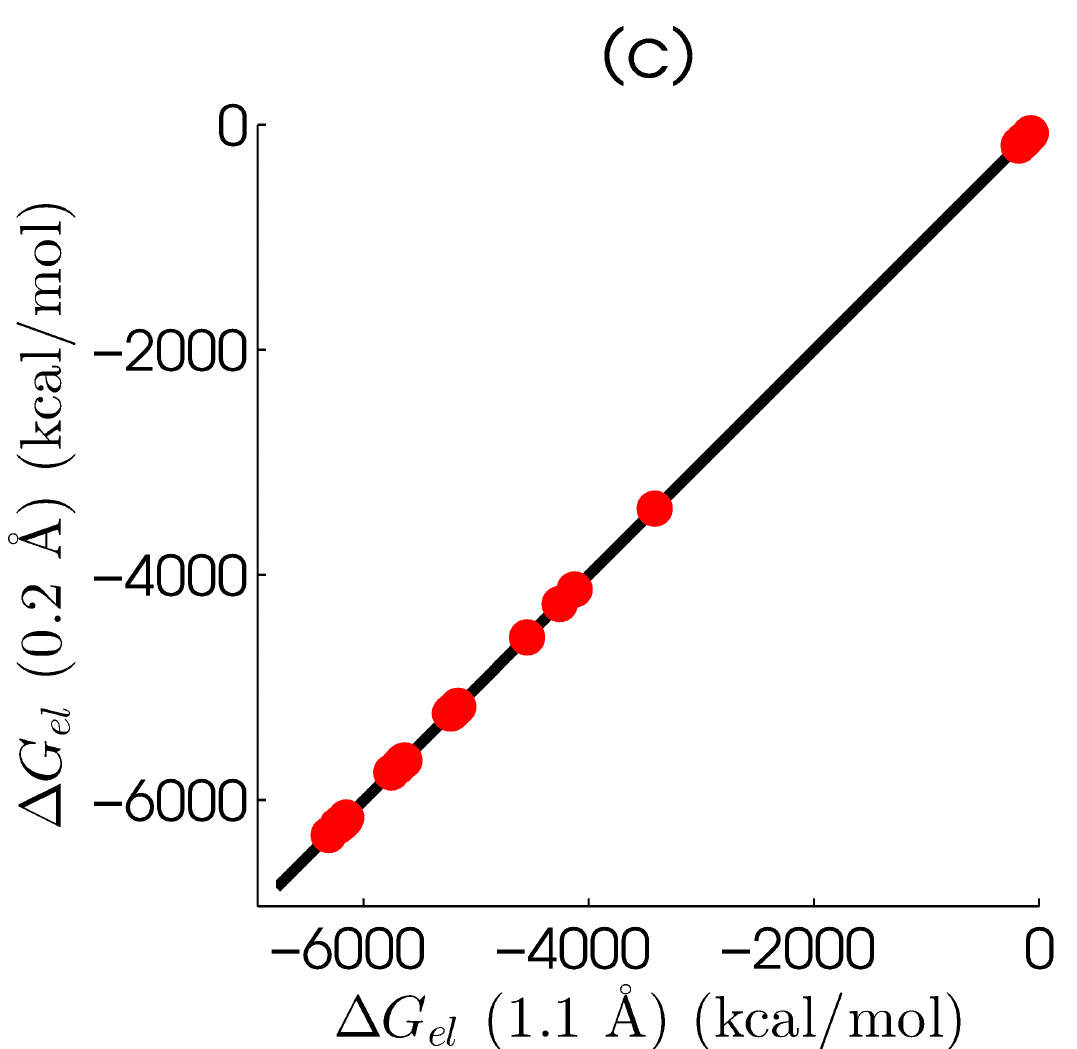}\\
		\vspace*{0.5cm}
		\includegraphics[width=0.30\columnwidth]{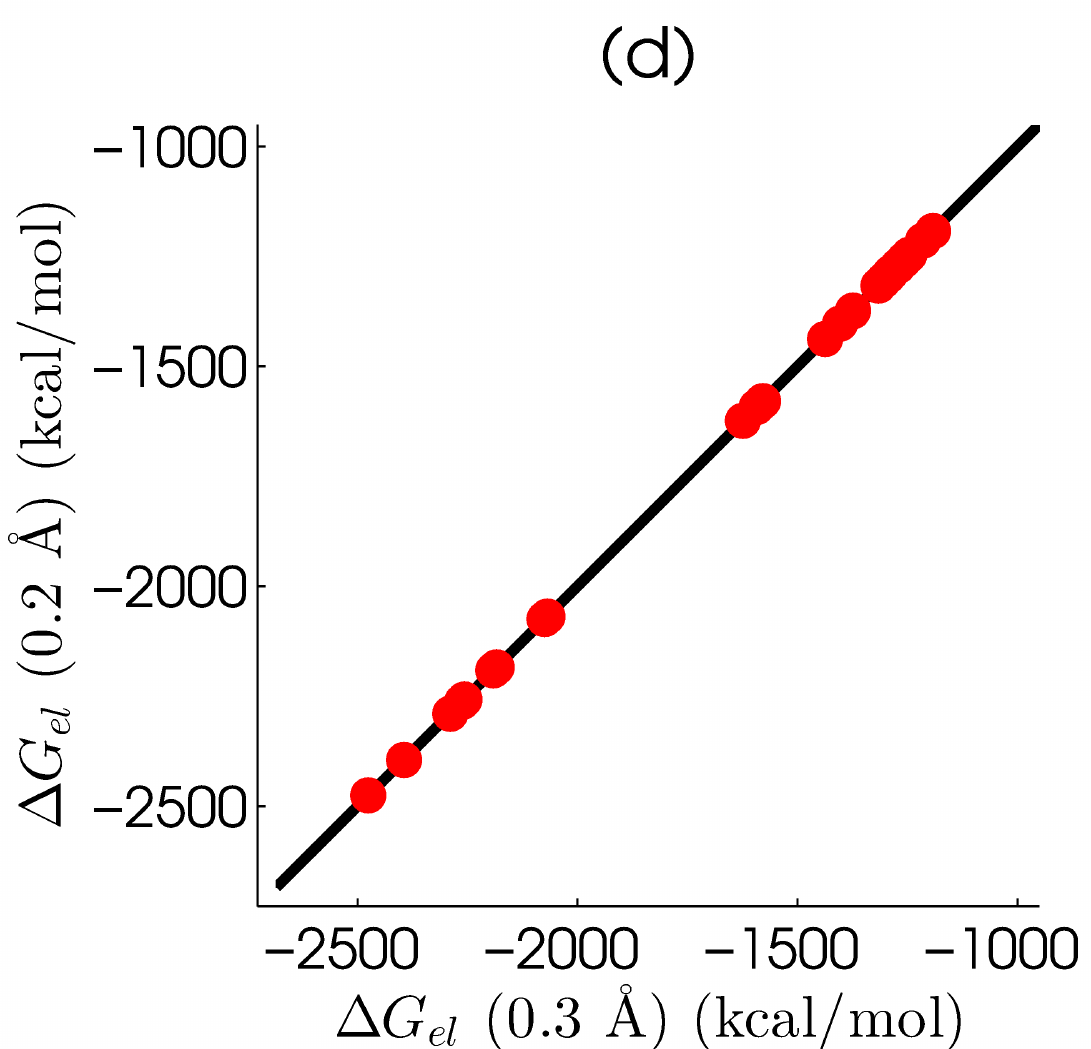}\quad
		\includegraphics[width=0.30\columnwidth]{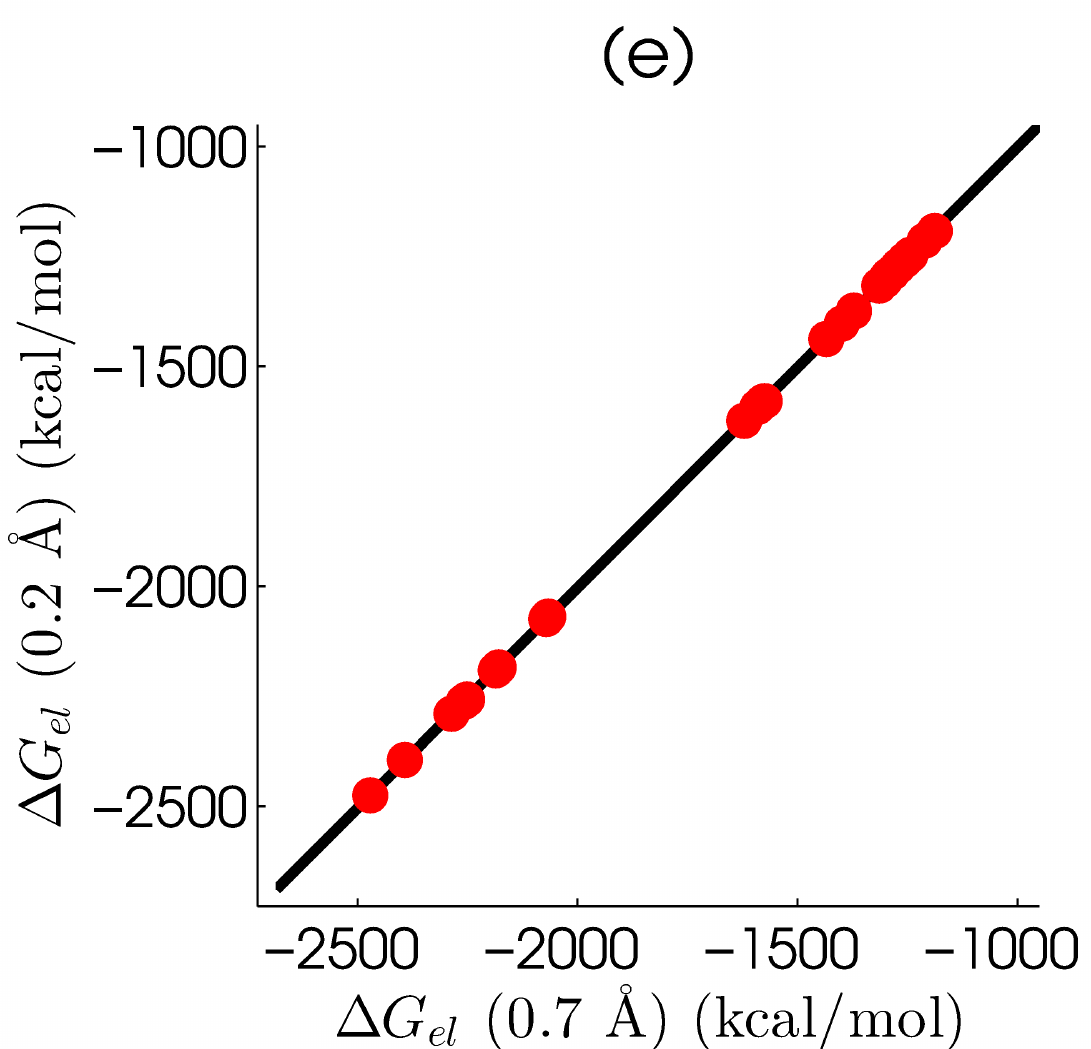}\quad
		\includegraphics[width=0.30\columnwidth]{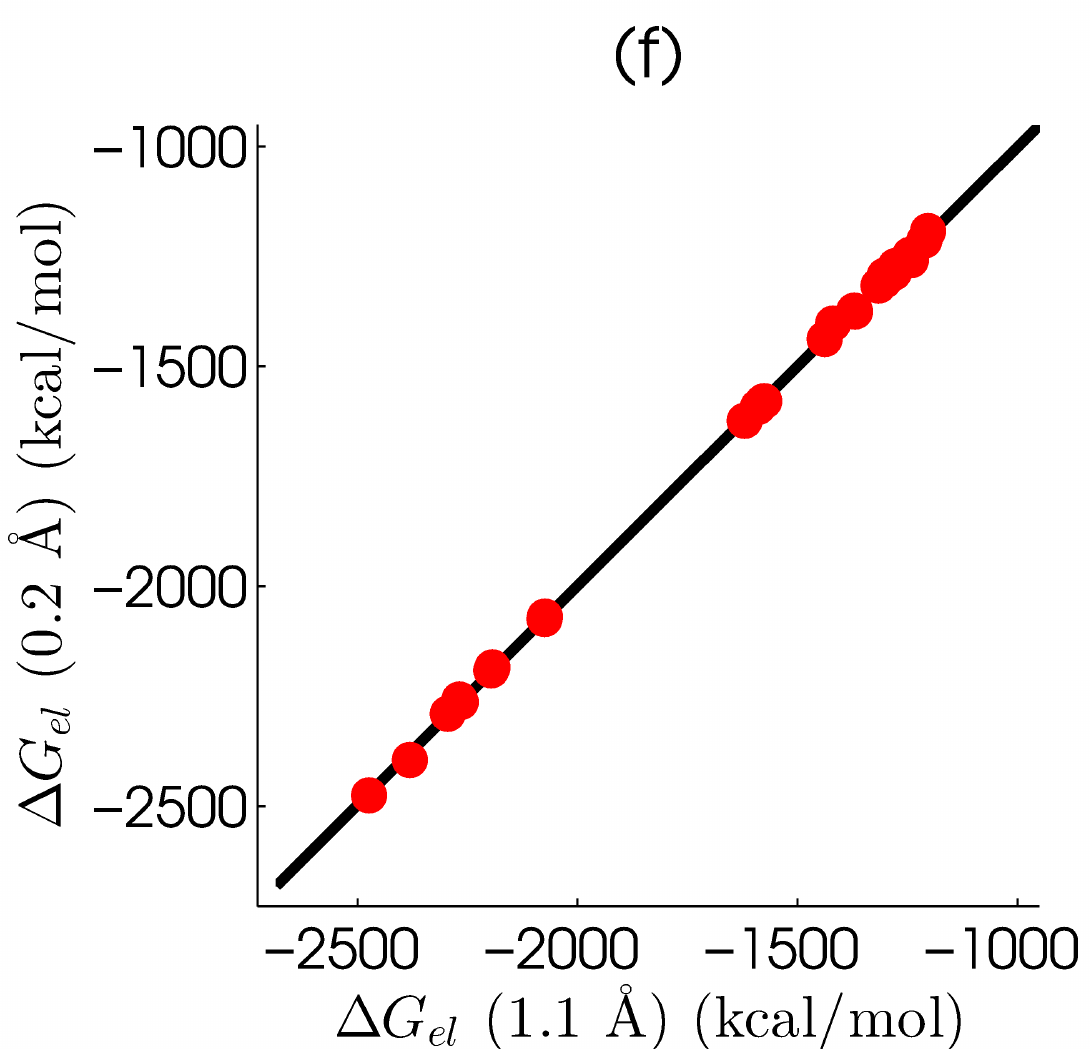}\\
		\vspace*{0.5cm}
		\includegraphics[width=0.30\columnwidth]{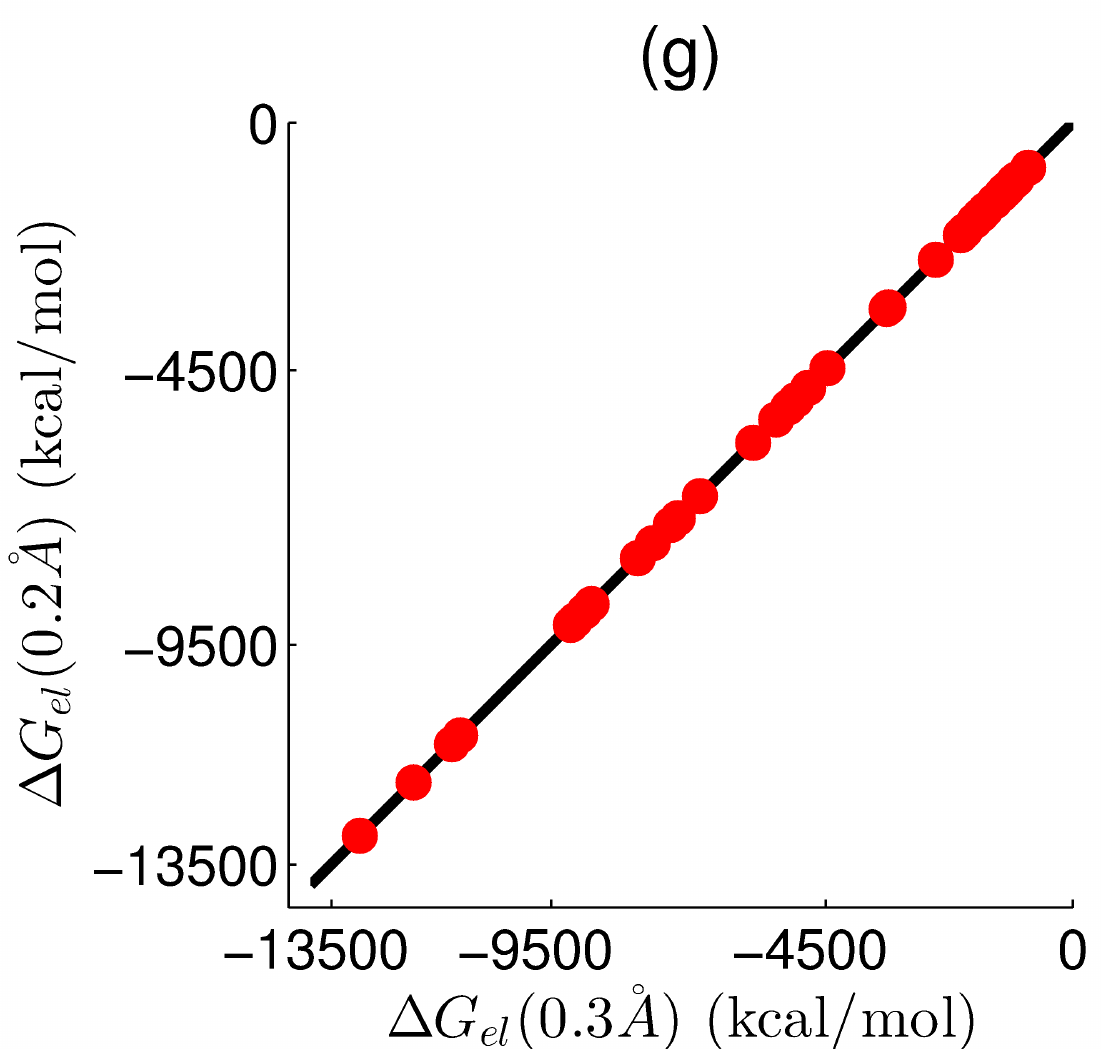}\quad
		\includegraphics[width=0.30\columnwidth]{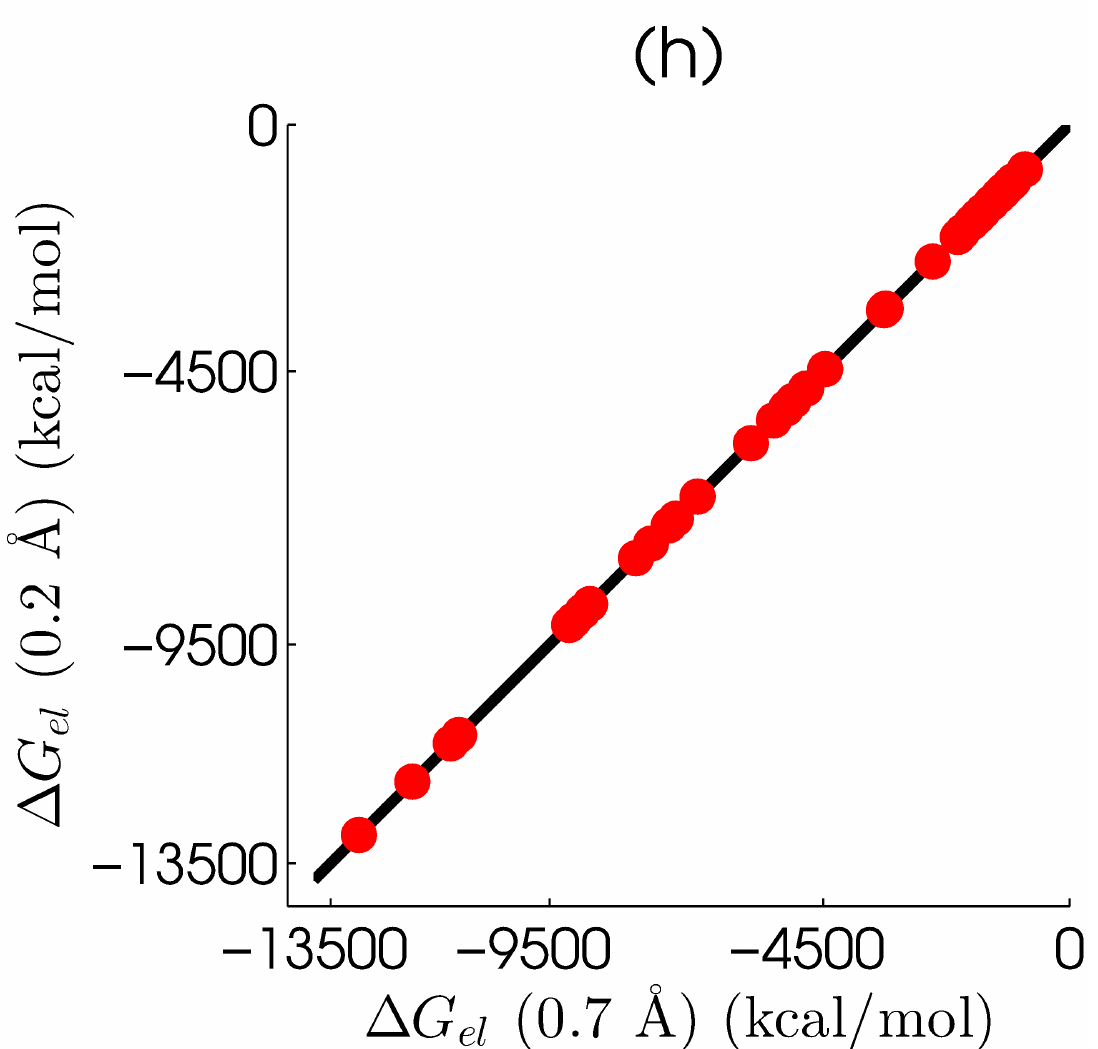}\quad
		\includegraphics[width=0.30\columnwidth]{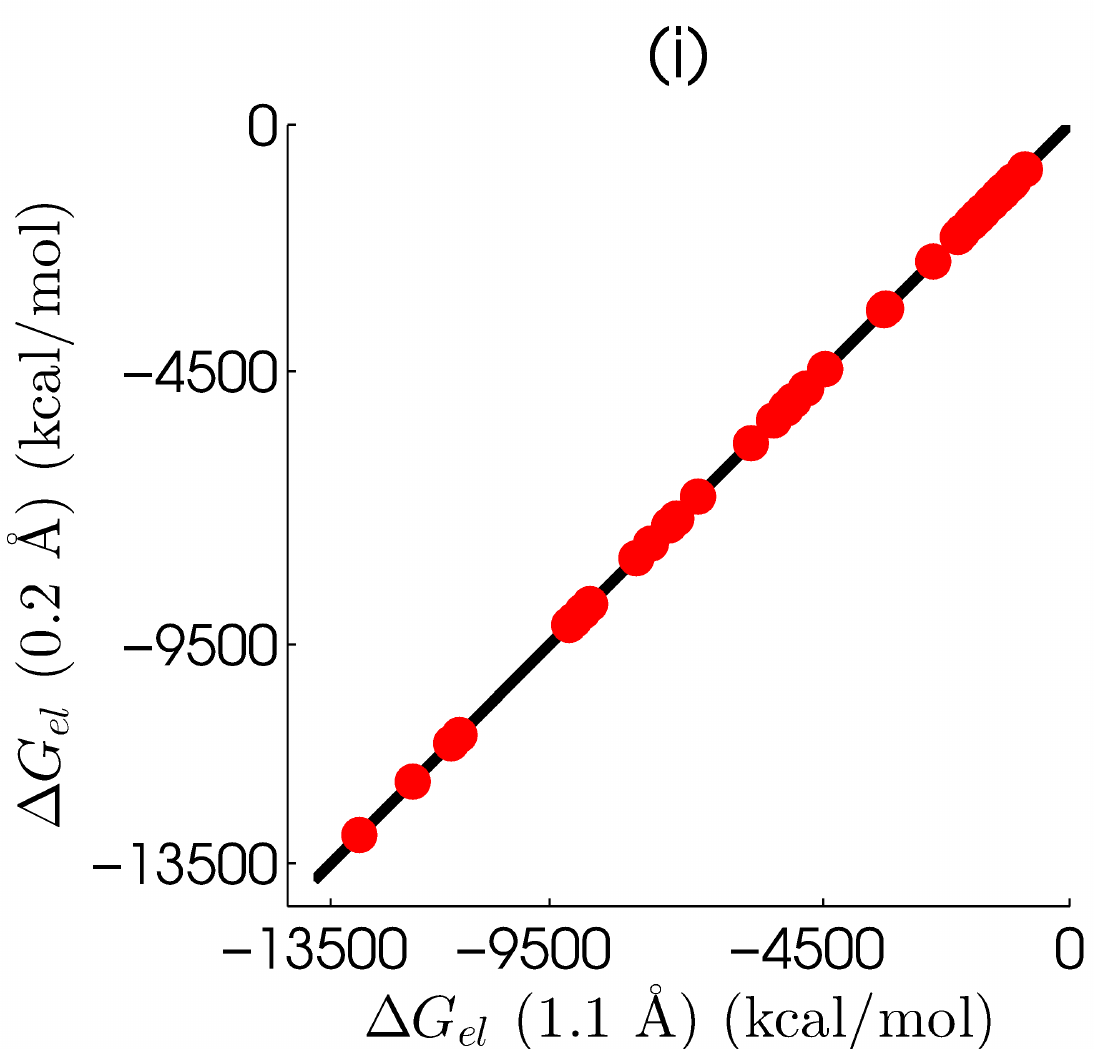}
		\caption{Electrostatic solvation free energy, for all complexes and unbounded components of three data sets, with different grid sizes plotted against the one computed with a finest grid size of $h=\SI{0.2}{\angstrom}$. (a) DNA-drug with pair ($\SI{0.2}{\angstrom}$,$\SI{0.3}{\angstrom}$); (b) DNA-drug with pair ($\SI{0.2}{\angstrom}$,$\SI{0.7}{\angstrom}$); (c) DNA-drug with pair ($\SI{0.2}{\angstrom}$,$\SI{1.1}{\angstrom}$); (d) Barnase-barstar with pair ($\SI{0.2}{\angstrom}$,$\SI{0.3}{\angstrom}$); (e) Barnase-barstar with pair ($\SI{0.2}{\angstrom}$,$\SI{0.7}{\angstrom}$); (f) Barnase-Barstar with pair ($\SI{0.2}{\angstrom}$,$\SI{1.1}{\angstrom}$); (g) RNA-peptide with pair ($\SI{0.2}{\angstrom}$,$\SI{0.3}{\angstrom}$); (h) RNA-peptide with pair ($\SI{0.2}{\angstrom}$,$\SI{0.7}{\angstrom}$); (i) RNA-peptide with pair ($\SI{0.2}{\angstrom}$,$\SI{1.1}{\angstrom}$). }
		\label{fig.ele}
	\end{center}
\end{figure} 

\begin{figure}[!tb]
	\begin{center}
		\includegraphics[width=0.70\columnwidth]{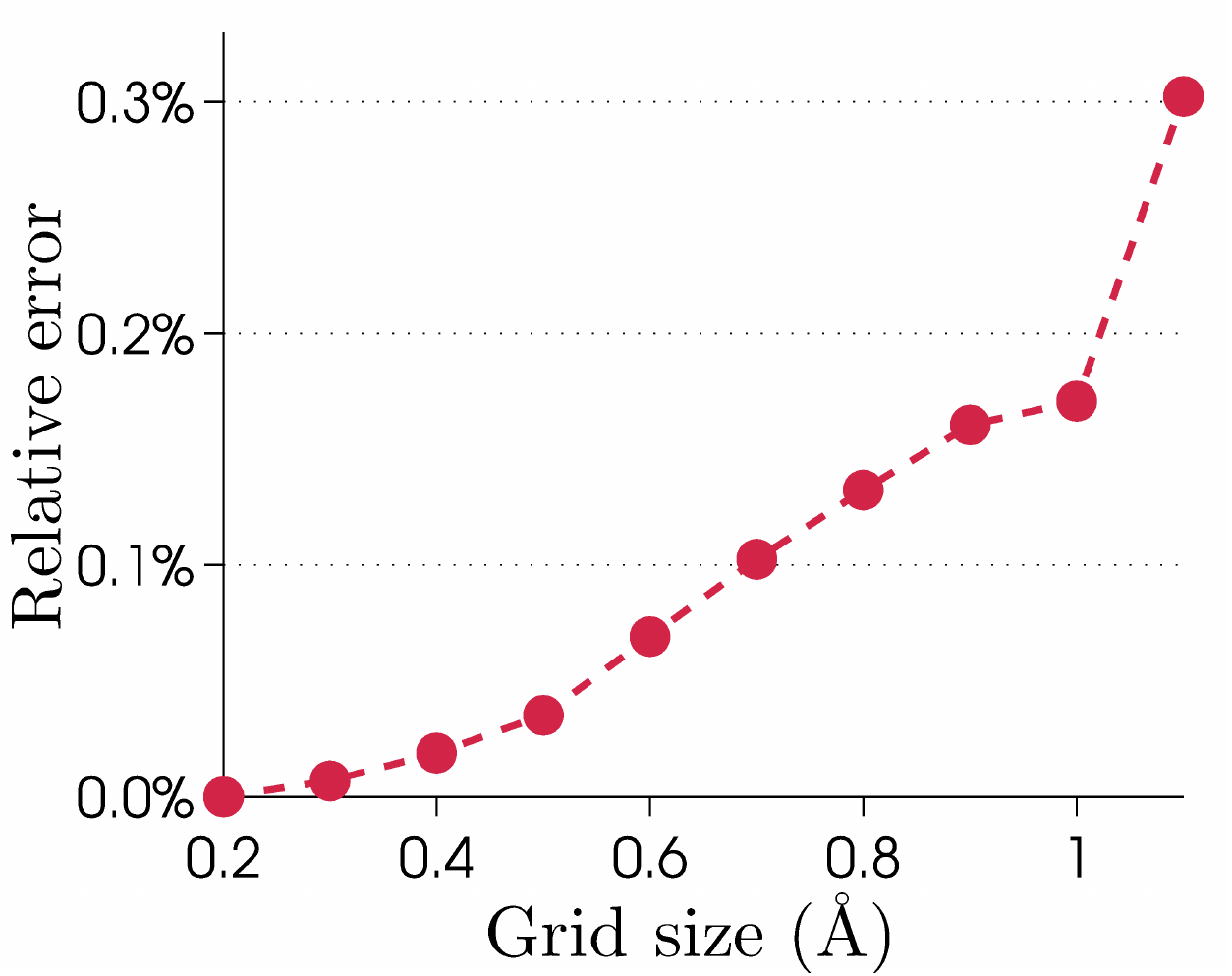}
		\caption{Averaged relative absolute error of the electrostatic solvation free energies for all the 153 molecules with mesh size refinements from $\SI{1.1}{\angstrom}$ to $\SI{0.2}{\angstrom}$.}
		\label{fig.rel_err}
	\end{center}
\end{figure}

\begin{table}[!tb]
\centering
\caption{$R^2$ values and best fitting lines of electrostatic solvation free energies with different grid sizes.}
\label{tab.r2_elec}
\begin{tabular}{lcll}
\hline
                & Grid sizes (pair) & \multicolumn{1}{c}{$R^2$} & \multicolumn{1}{c}{Best fitting line} \\ \hline
DNA-drug        & (0.2,0.3)         & 1.0000                    & $y=1.0000x-0.0196$                    \\
                & (0.2,0.4)         & 1.0000                    & $y=1.0000x-0.0081$                    \\
                & (0.2,0.5)         & 1.0000                    & $y=1.0001x-0.0621$                    \\
                & (0.2,0.6)         & 1.0000                    & $y=1.0001x-0.2230$                    \\
                & (0.2,0.7)         & 1.0000                    & $y=1.0003x-0.2537$                    \\
                & (0.2,0.8)         & 1.0000                    & $y=1.0003x-0.4161$                    \\
                & (0.2,0.9)         & 1.0000                    & $y=1.0003x-0.2999$                    \\
                & (0.2,1.0)         & 1.0000                    & $y=1.0005x-0.0066$                    \\
                & (0.2,1.1)         & 1.0000                    & $y=1.0004x-0.2485$                    \\
                \\
Barnase-barstar & (0.2,0.3)         & 1.0000                    & $y=1.0002x+0.1590$                    \\
                & (0.2,0.4)         & 1.0000                    & $y=1.0005x+0.3524$                    \\
                & (0.2,0.5)         & 1.0000                    & $y=1.0012x+0.8735$                    \\
                & (0.2,0.6)         & 1.0000                    & $y=1.0010x-0.2246$                    \\
                & (0.2,0.7)         & 1.0000                    & $y=1.0017x-0.3748$                    \\
                & (0.2,0.8)         & 1.0000                    & $y=1.0009x-0.9576$                    \\
                & (0.2,0.9)         & 0.9999                    & $y=1.0015x+0.4749$                    \\
                & (0.2,1.0)         & 0.9999                    & $y=0.9986x-2.9739$                    \\
                & (0.2,1.1)         & 0.9997                    & $y=0.9972x-4.3801$                    \\
                \\
RNA-peptide     & (0.2,0.3)         & 1.0000                    & $y=1.0000x-0.0445$                    \\
                & (0.2,0.4)         & 1.0000                    & $y=1.0000x-0.1333$                    \\
                & (0.2,0.5)         & 1.0000                    & $y=1.0000x-0.3343$                    \\
                & (0.2,0.6)         & 1.0000                    & $y=1.0000x-0.1916$                    \\
                & (0.2,0.7)         & 1.0000                    & $y=1.0001x-0.5377$                    \\
                & (0.2,0.8)         & 1.0000                    & $y=1.0001x-0.8198$                    \\
                & (0.2,0.9)         & 1.0000                    & $y=1.0002x-0.9564$                    \\
                & (0.2,1.0)         & 1.0000                    & $y=1.0003x-0.8868$                    \\
                & (0.2,1.1)         & 1.0000                    & $y=1.0005x-2.2504$                    \\
\hline
\end{tabular}
\end{table}

We first examine the  accuracy and robustness of our MIBPB solver in predicting the electrostatic solvation free energies of the aforementioned three data sets. Some previous literature   \cite{Feig:2004a,reyes2000structure} has recognized that a grid size of $h=\SI{0.5}{\angstrom}$ is small enough to produce a reliable $\Delta G_{\text{el}}$. Such an observation certainly remains for the MIBPB solver. In fact, our PB solver is able to deliver a very well-convergent calculations of  electrostatic solvation free energies at as coarse grid sizes as $\SI{1.0}{\angstrom}$ and $\SI{1.1}{\angstrom}$. 

In the current calculations, the finest grid size is chosen to be $\SI{0.2}{\angstrom}$, and the coarser grid sizes are between $\SI{0.3}{\angstrom}$ and $\SI{1.1}{\angstrom}$. Figure \ref{fig.ele} depicts the correlations of  $\Delta G_{\text{el}}$ at various meshes for all complexes and unbounded components of three data sets. The electrostatic solvation free energies obtained  at the finest grid spacing of $\SI{0.2}{\angstrom}$ are plotted against those computed from  coarser grid spacings of  $\SI{0.3}{\angstrom}$, $\SI{0.7}{\angstrom}$ and $\SI{1.1}{\angstrom}$. Obviously, the best fitting lines for these data at various coarse grid spacings produce near perfect alignments between the finest mesh results and those from coarse meshes.  As shown in Table \ref{tab.r2_elec}, $R^2$ and slope values at the pair of grid sizes $(\SI{0.2}{\angstrom},\SI{1.1}{\angstrom})$ for DNA-drug, barnase-barstar and RNA-peptide are, respectively, $(1.0000,1.0004)$, $(0.9997,0.9972)$, and $(1.0000,1.0005)$. These results indicate the accuracy and robustness in the MIBPB prediction  of electrostatic solvation free energies ($\Delta G_{\text{el}}$). Table S1, in the Supporting Information, reports the values $\Delta G_{\text{el}}$ for all the 51 complexes and associated 102 components studied in this work. Finally, we  examine the performance of our solver by considering the relative absolute error, the difference between results obtained with coarser and the finest grid spacings, defined as follows
\begin{equation}\label{eq.rel_err}
\text{Relative absolute error}\doteq \left|\frac{\Delta G_{\text{el},h}-\Delta G_{\text{el,h=0.2}}}{\Delta G_{\text{el,h=0.2}}}\right|.
\end{equation}

Figure \ref{fig.rel_err} illustrates  the averaged relative absolute errors, i.e., the average of relative absolute errors designated in Eq. \eqref{eq.rel_err} over all the 153 discussed molecules, at different mesh sizes. It can be seen from Fig. \ref{fig.rel_err} that the averaged relative absolute errors at all studied cases are less than $0.31 \%$, and for any grid spacing smaller than $\SI{1.1}{\angstrom}$, these errors are always below $0.2\%$. This behavior further indicates the grid size independence of our PB solver over the normal grid-size range in molecular biophysical applications.

\subsection{The influence of grid spacing in  $\Delta\Delta G_{\text{el}}$   estimation}

Motivated by well-converged estimations of electrostatic solvation free energies at very coarse grid spacings as previously discussed, we are interested in predicting the binding free energies for all RNA-drug, barnase-barstar, and RNA-peptide complexes using our MIBPB package.

\begin{figure}[!tb]
	\begin{center}
		\includegraphics[width=0.30\columnwidth]{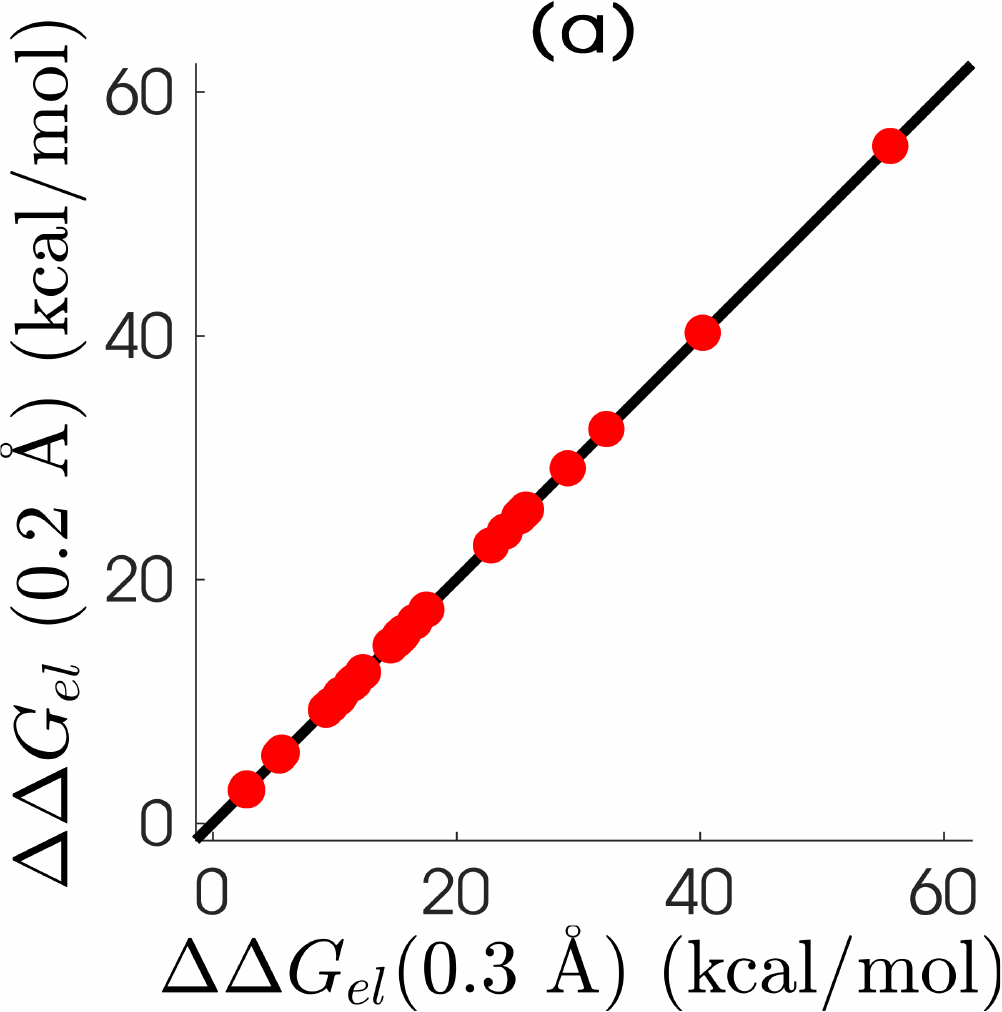}\quad
		\includegraphics[width=0.30\columnwidth]{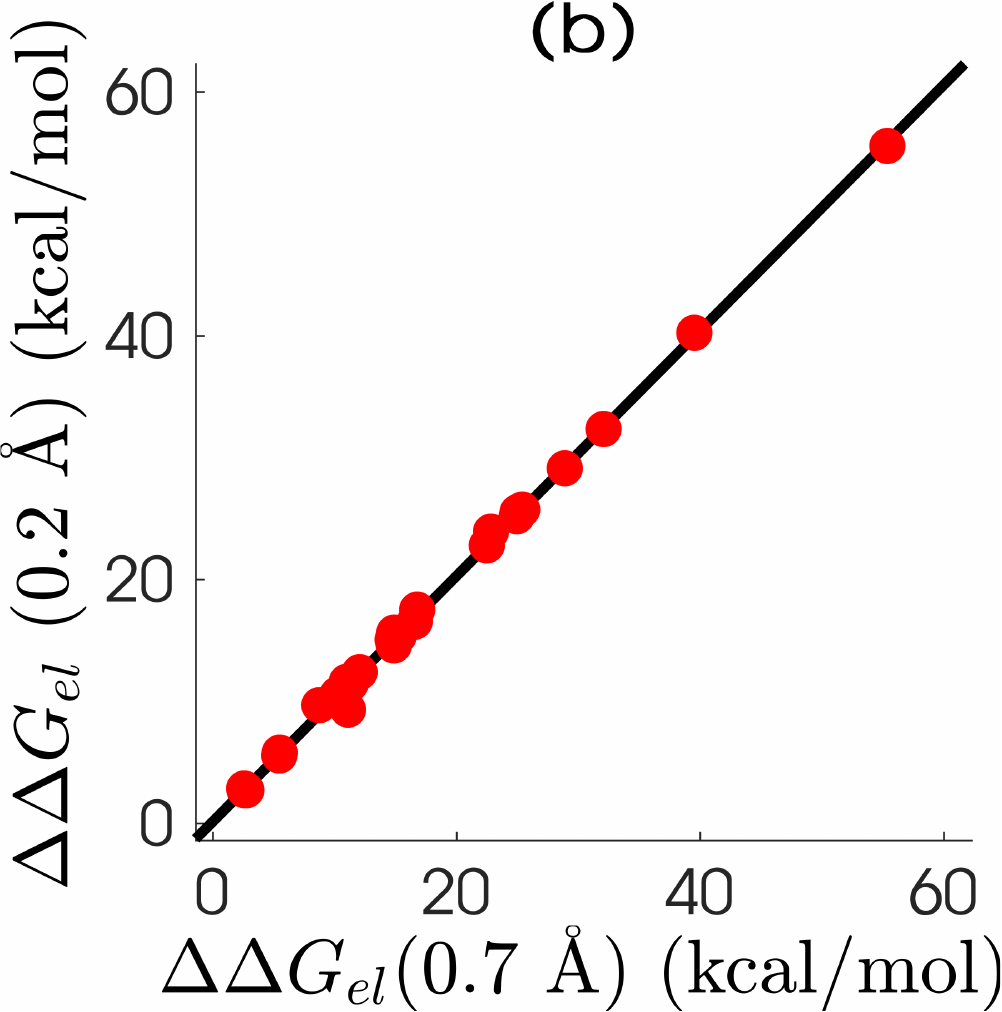}\quad
		\includegraphics[width=0.30\columnwidth]{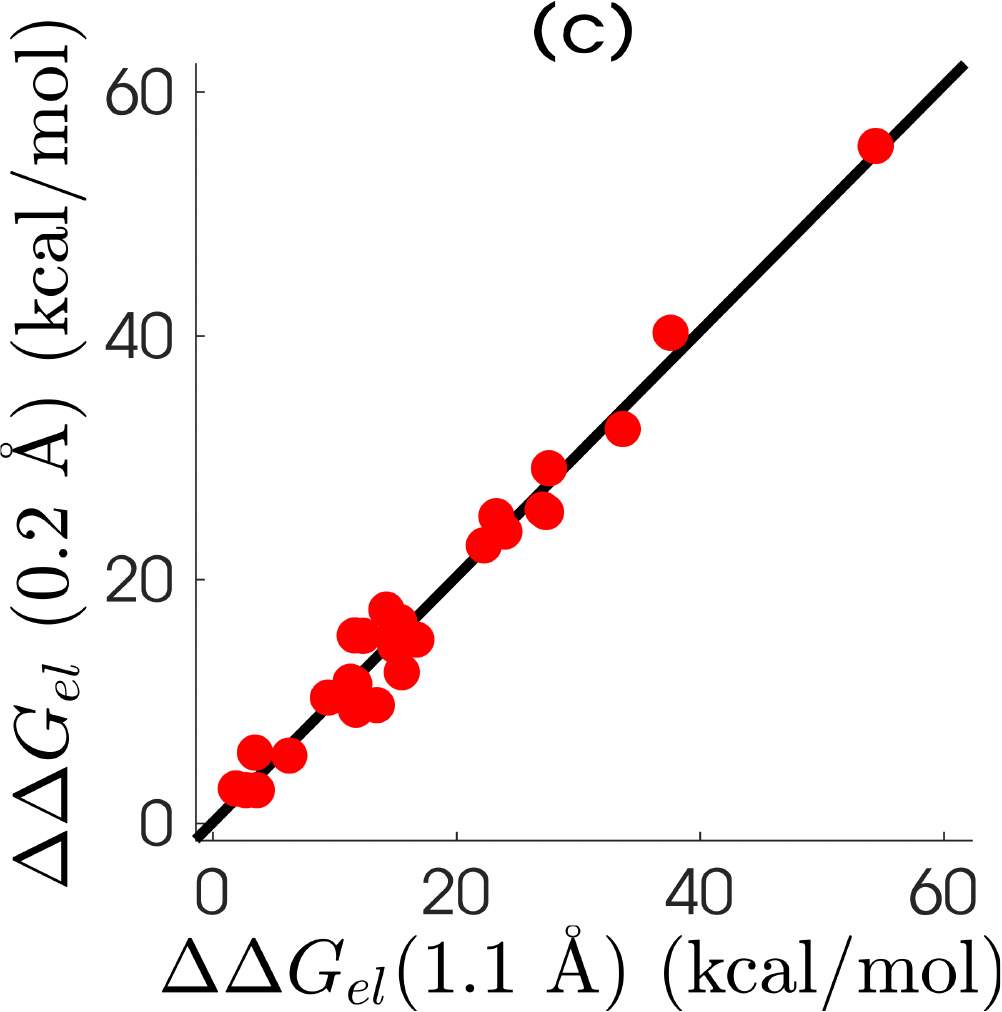}\\
		\vspace*{0.5cm}
		\includegraphics[width=0.30\columnwidth]{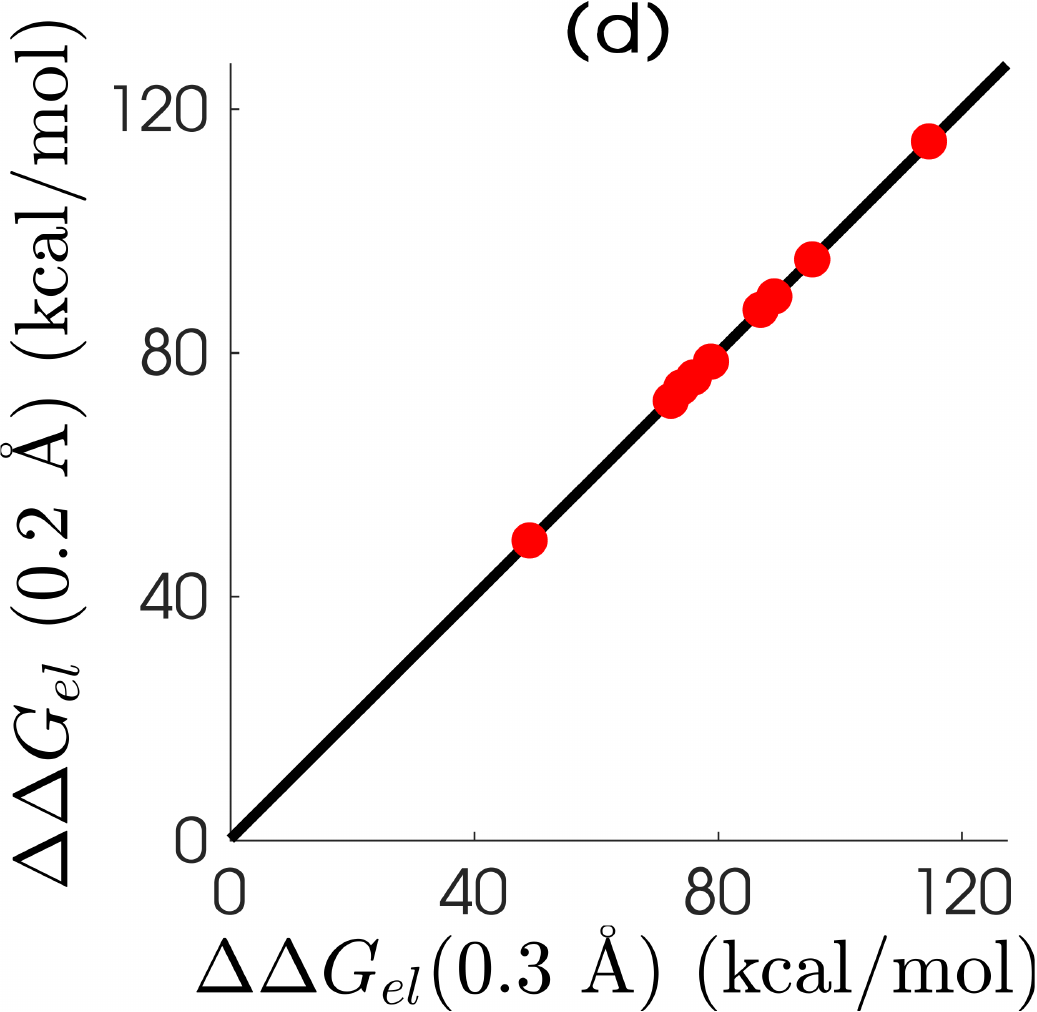}\quad
		\includegraphics[width=0.30\columnwidth]{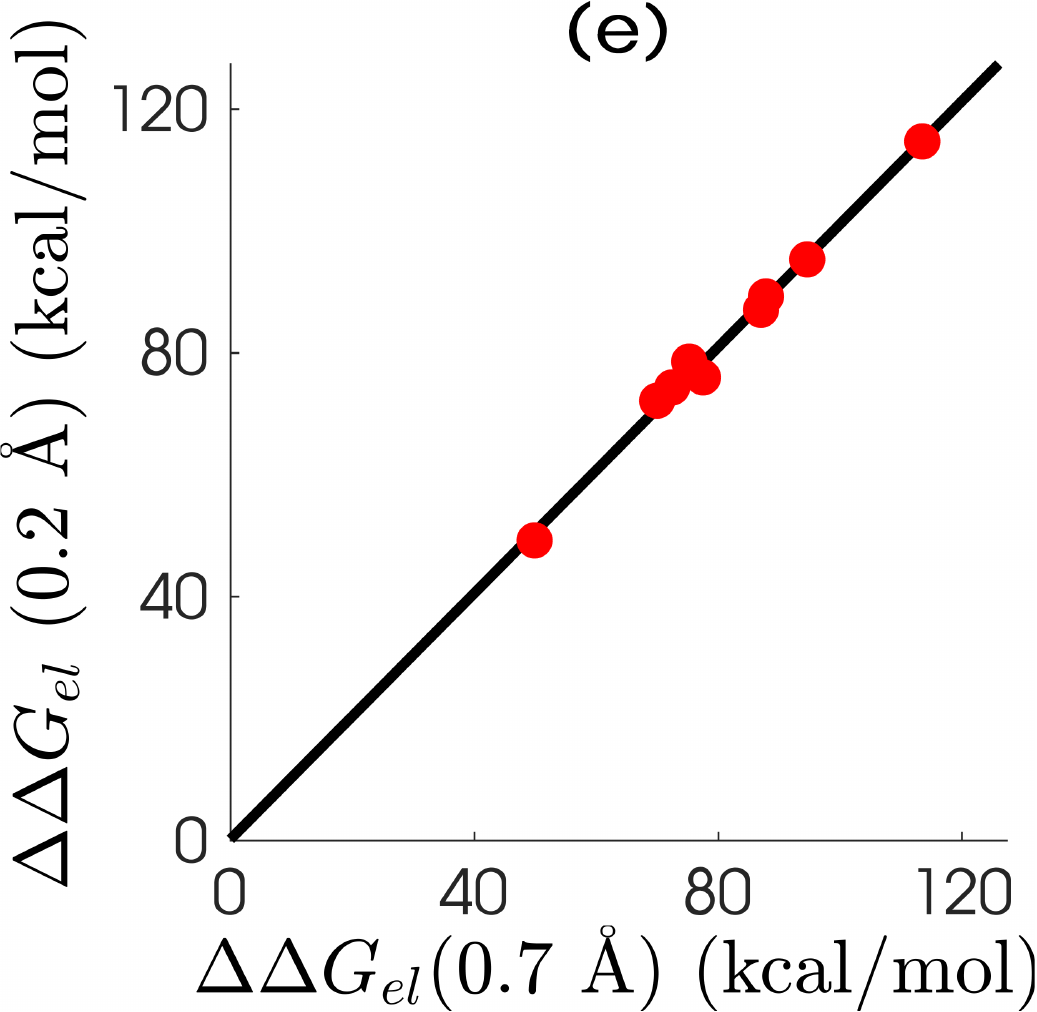}\quad
		\includegraphics[width=0.30\columnwidth]{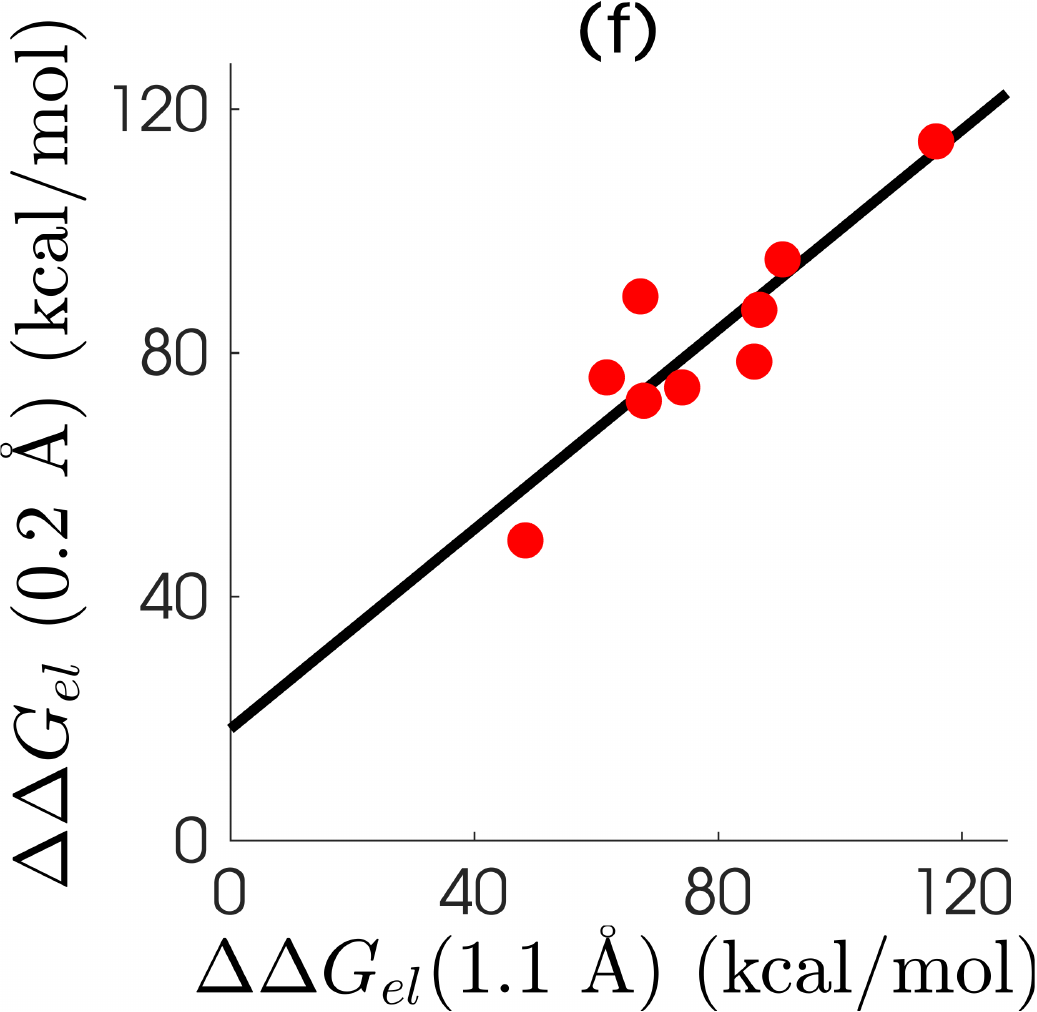}\\
		\vspace*{0.5cm}
		\includegraphics[width=0.30\columnwidth]{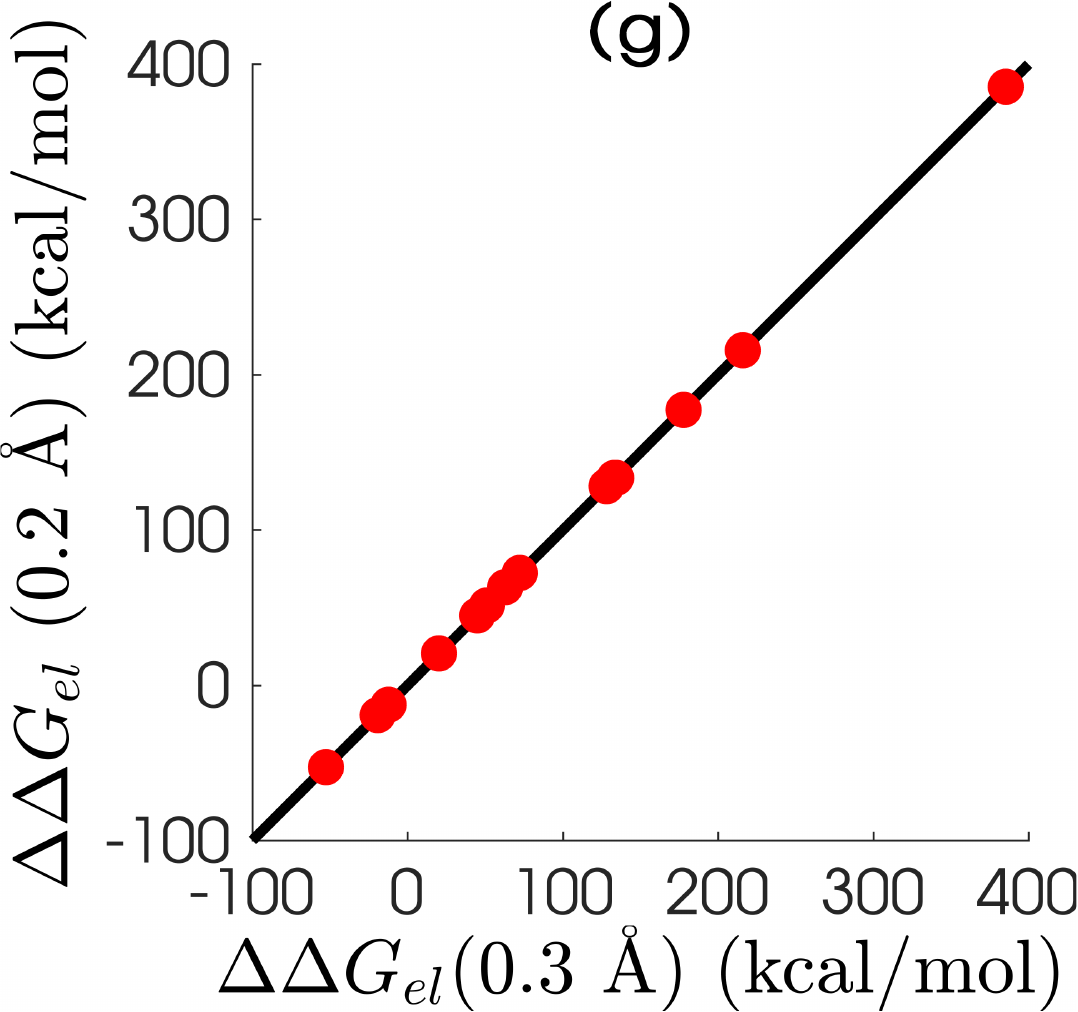}\quad
		\includegraphics[width=0.30\columnwidth]{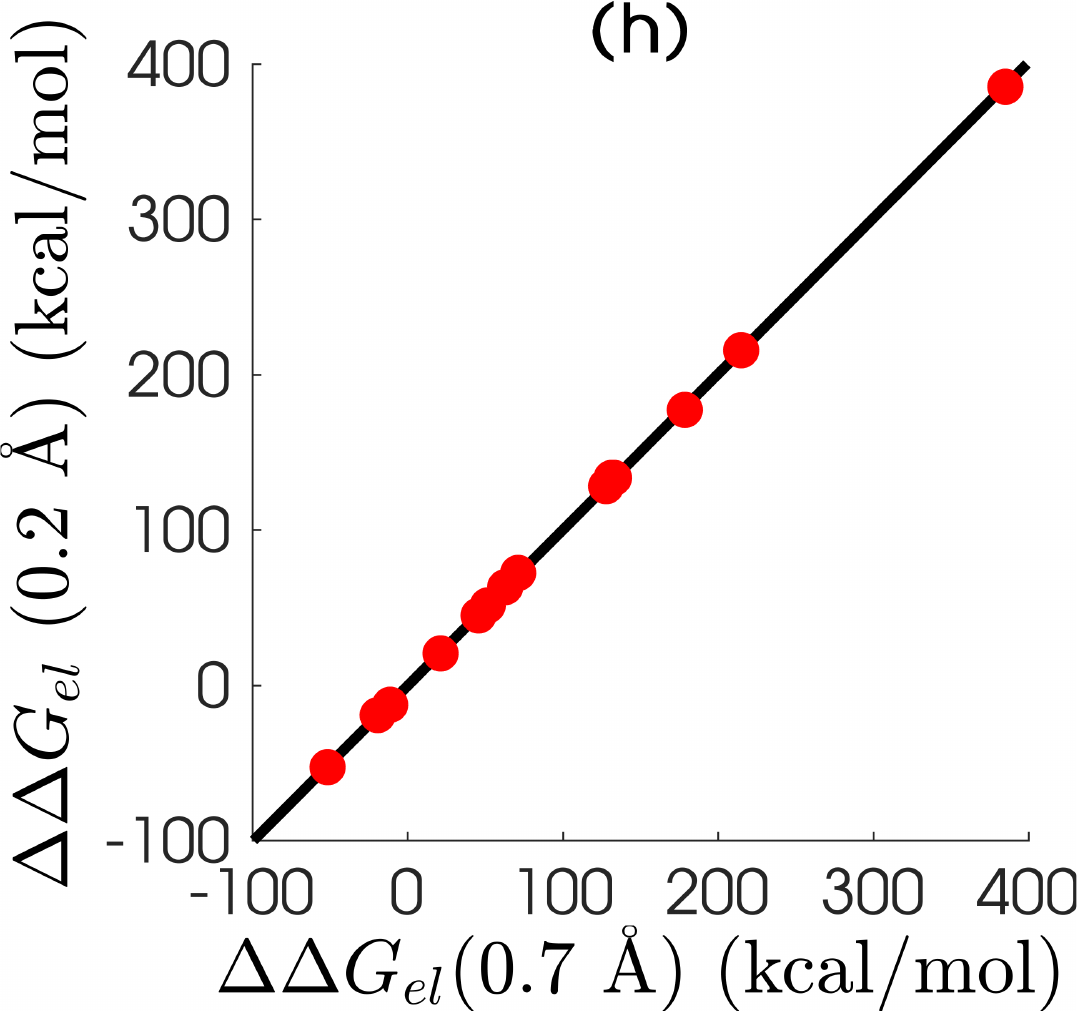}\quad
		\includegraphics[width=0.30\columnwidth]{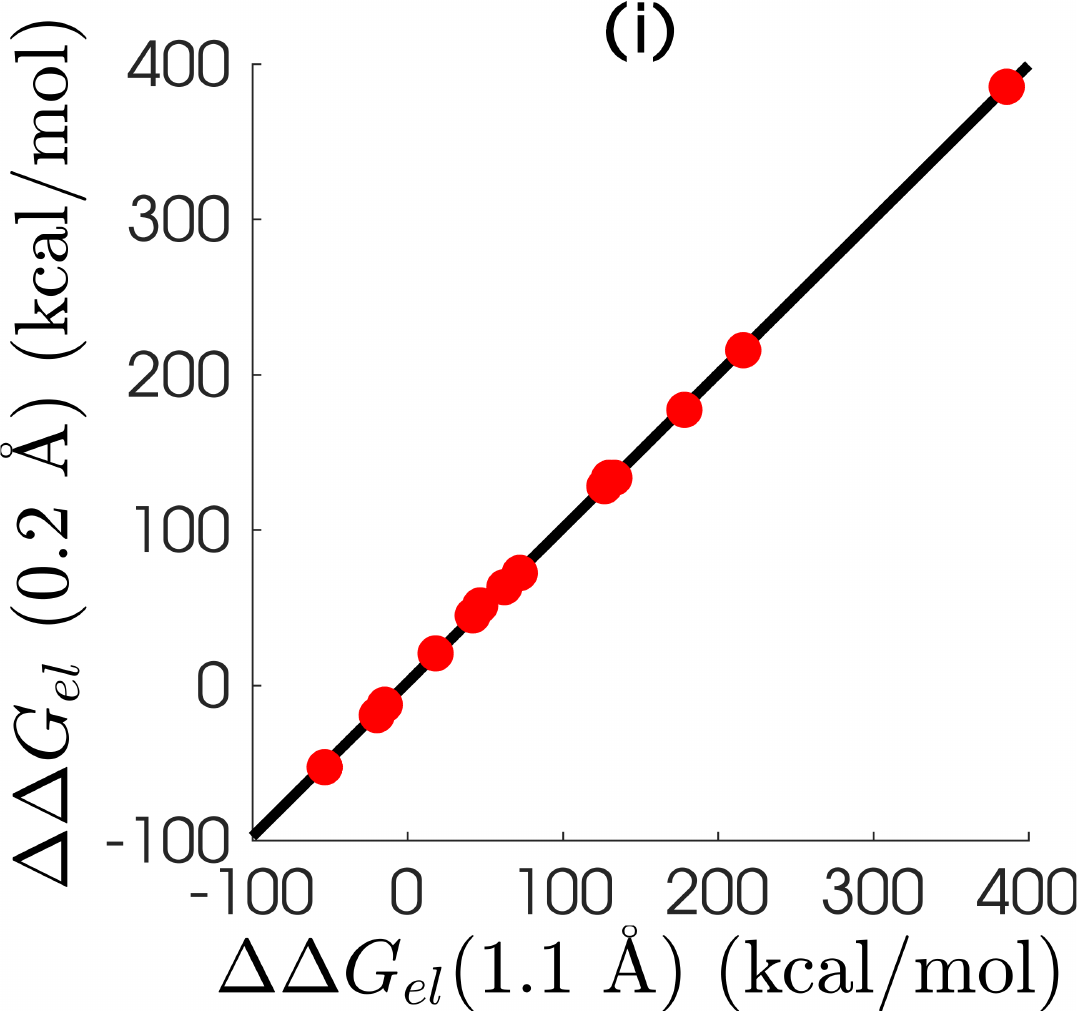}
		\caption{Electrostatic binding free energy, for all complexes  with different grid sizes plotted against the one computed with a finest grid size of $h=\SI{0.2}{\angstrom}$. (a) DNA-drug with pair ($\SI{0.2}{\angstrom}$,$\SI{0.3}{\angstrom}$); (b) DNA-drug with pair ($\SI{0.2}{\angstrom}$,$\SI{0.7}{\angstrom}$); (c) DNA-drug with pair ($\SI{0.2}{\angstrom}$,$\SI{1.1}{\angstrom}$); (d) Barnase-barstar with pair ($\SI{0.2}{\angstrom}$,$\SI{0.3}{\angstrom}$); (e) Barnase-barstar with pair ($\SI{0.2}{\angstrom}$,$\SI{0.7}{\angstrom}$); (f) Barnase-barstar with pair ($\SI{0.2}{\angstrom}$,$\SI{1.1}{\angstrom}$); (g) RNA-peptide with pair ($\SI{0.2}{\angstrom}$,$\SI{0.3}{\angstrom}$); (h) RNA-peptide with pair ($\SI{0.2}{\angstrom}$,$\SI{0.7}{\angstrom}$); (i) RNA-peptide with pair ($\SI{0.2}{\angstrom}$,$\SI{1.1}{\angstrom}$). }
		\label{fig.bd}
	\end{center}
\end{figure} 

\begin{table}[!tb]
\centering
\caption{$R^2$ values and best fitting lines of electrostatic binding free energies with different grid sizes.}
\label{tab.r2_bd}
\begin{tabular}{lcll}
\hline
                & Grid sizes (pair) & \multicolumn{1}{c}{$R^2$} & \multicolumn{1}{c}{Best fitting line} \\ \hline
DNA-drug        & (0.2,0.3)         & 1.0000                    &     $y=0.9993x   +  0.0194$                   \\
				& (0.2,0.4)         & 0.9999                    &     $y=0.9987x   +  0.0273$                   \\
				& (0.2,0.5)         & 0.9998                    &     $y=1.0028x   +  0.0164$                   \\
				& (0.2,0.6)         & 0.9991                    &     $y=1.0047x   +  0.2256$                   \\
				& (0.2,0.7)         & 0.9982                    &     $y=1.0074x   +  0.1394$                   \\
				& (0.2,0.8)         & 0.9966                    &     $y=1.0110x   +  0.1484$                   \\
				& (0.2,0.9)         & 0.9906                    &     $y=0.9655x   +  1.2385$                   \\
				& (0.2,1.0)         & 0.9875                    &     $y=0.9827x   +  0.5894$                   \\
				& (0.2,1.1)         & 0.9747                    &     $y=1.0081x   +  0.0709$                   \\
                \\
Barnase-barstar & (0.2,0.3)   &   0.9999   &   $y=0.9974x + 0.2035$ \\  
				& (0.2,0.4)   &   0.9995   &   $y=0.9997x - 0.0492$\\
				& (0.2,0.5)   &   0.9923   &   $y=1.0318x - 2.7755$\\
				& (0.2,0.6)   &   0.9946   &   $y=0.9878x + 1.5525$\\
				& (0.2,0.7)   &   0.9932   &   $y=1.0090x + 0.1819$\\
				& (0.2,0.8)   &   0.9883   &   $y=0.9766x + 3.7333$\\
				& (0.2,0.9)   &   0.9493   &   $y=0.9382x + 5.3970$\\
				& (0.2,1.0)   &   0.9384   &   $y=1.0912x - 3.8377$\\
				& (0.2,1.1)   &   0.8002   &   $y=0.8187x + 18.2837$\\  
                \\
RNA-peptide     & (0.2,0.3)  &  1.0000  &    $y=0.9997x - 0.0655$ \\
				& (0.2,0.4)  &  1.0000  &    $y=1.0001x - 0.1106$ \\
				& (0.2,0.5)  &  1.0000  &    $y=1.0012x - 0.2755$ \\
				& (0.2,0.6)  &  1.0000  &    $y=0.9999x + 0.2021$ \\
				& (0.2,0.7)  &  0.9999  &    $y=1.0037x - 0.3756$ \\
				& (0.2,0.8)  &  1.0000  &    $y=1.0004x + 0.6673$ \\
				& (0.2,0.9)  &  0.9999  &    $y=0.9927x + 1.9755$ \\
				& (0.2,1.0)  &  0.9997  &    $y=0.9923x + 2.8775$ \\
				& (0.2,1.1)  &  0.9998  &    $y=0.9937x + 1.7992$ \\
\hline
\end{tabular}
\end{table}

\begin{table}[!tb]
\begin{threeparttable}[b]
\scriptsize
\centering
\caption{Electrostatic binding free energies (in units of kcal/mol), $\Delta\Delta G_{\text{el}}$, for all of the complexes used in this study at different grid sizes.}
\label{tab.binding_eng}
\begin{tabular}{@{}llrrrrrrrrrr@{}}
\toprule
 & complexes & $\SI{1.1}{\angstrom}$ & $\SI{1.0}{\angstrom}$ & $\SI{0.9}{\angstrom}$ & $\SI{0.8}{\angstrom}$ & $\SI{0.7}{\angstrom}$ & $\SI{0.6}{\angstrom}$ & $\SI{0.5}{\angstrom}$ & $\SI{0.4}{\angstrom}$ & $\SI{0.3}{\angstrom}$ & $\SI{0.2}{\angstrom}$ \\ \midrule
DNA-drug    & 102d      	&   9.45        &  10.50        &   8.73         & 10.01        &  10.21        &  10.76        &  10.53        &  10.45        &  10.34        &  10.31 \\
			& 109d          &   3.61        &   2.18        &   2.30         &  3.72        &   2.63        &   2.07        &   2.66        &   2.69        &   2.82        &   2.72 \\
			& 121d          &  23.95        &  23.99        &  23.80         & 24.05        &  22.84        &  23.65        &  24.10        &  23.94        &  23.96        &  23.93 \\
			& 127d          &  27.60        &  28.80        &  28.45         & 28.89        &  28.88        &  28.93        &  29.27        &  29.16        &  29.12        &  29.12 \\
			& 129d          &  37.58        &  40.23        &  39.92         & 39.03        &  39.52        &  39.90        &  40.04        &  40.15        &  40.20        &  40.24 \\
			& 166d          &  15.04        &  16.60        &  13.97         & 14.93        &  14.91        &  15.49        &  15.47        &  15.62        &  15.67        &  15.67 \\
			& 195d          &   2.74        &   3.73        &   2.80         &  2.63        &   2.77        &   2.77        &   2.63        &   2.69        &   2.72        &   2.73 \\
			& 1d30          &  11.27        &  10.01        &  10.71         &  9.31        &  10.26        &  10.12        &  10.40        &  10.75        &  10.53        &  10.59 \\
			& 1d63          &  15.51        &  12.83        &   7.08         & 12.56        &  12.07        &  11.24        &  12.10        &  12.29        &  12.34        &  12.39 \\
			& 1d64          &  14.98        &  14.11        &  14.26         & 14.03        &  14.86        &  14.24        &  14.51        &  14.57        &  14.59        &  14.58 \\
			& 1d86          &  27.37        &  25.53        &  24.50         & 25.88        &  25.04        &  25.57        &  25.39        &  25.44        &  25.50        &  25.54 \\
			& 1dne          &  22.26        &  22.73        &  22.32         & 22.92        &  22.48        &  22.62        &  22.58        &  22.74        &  22.83        &  22.81 \\
			& 1eel          &  16.71        &  17.06        &  14.94         & 14.82        &  14.77        &  14.60        &  15.08        &  14.85        &  15.15        &  15.07 \\
			& 1fmq          &  12.35        &  13.40        &  14.28         & 14.72        &  15.27        &  15.09        &  15.30        &  15.36        &  15.36        &  15.37 \\
			& 1fms          &  27.08        &  26.14        &  25.17         & 24.52        &  25.41        &  25.93        &  25.82        &  25.75        &  25.71        &  25.74 \\
			& 1jtl          &  11.62        &  10.99        &  11.80         & 11.47        &  11.30        &  11.28        &  11.37        &  11.28        &  11.41        &  11.45 \\
			& 1lex          &  13.47        &  10.37        &   9.79         &  9.44        &   8.74        &   9.74        &   9.81        &   9.70        &   9.70        &   9.70 \\
			& 1prp          &  11.30        &  11.93        &  10.78         & 10.88        &  11.01        &  11.55        &  11.55        &  11.49        &  11.61        &  11.61 \\
			& 227d          &   6.28        &   4.79        &   5.96         &  3.79        &   5.47        &   5.16        &   5.80        &   5.75        &   5.46        &   5.58 \\
			& 261d          &   1.91        &   3.00        &   1.75         &  1.55        &   2.60        &   2.79        &   2.97        &   2.76        &   2.80        &   2.85 \\
			& 264d          &  33.64        &  32.35        &  31.57         & 30.83        &  32.09        &  31.97        &  32.07        &  32.20        &  32.31        &  32.34 \\
			& 289d          &  15.32        &  15.71        &  17.94         & 16.70        &  16.57        &  16.21        &  16.22        &  16.56        &  16.59        &  16.56 \\
			& 298d          &  11.65        &  15.94        &  14.81         & 15.89        &  14.87        &  14.88        &  15.38        &  15.45        &  15.50        &  15.41 \\
			& 2dbe          &   3.48        &   5.06        &   3.49         &  6.14        &   5.51        &   5.77        &   5.76        &   5.88        &   5.68        &   5.81 \\
			& 302d          &  23.28        &  25.22        &  24.91         & 25.49        &  24.95        &  25.00        &  24.87        &  25.29        &  25.17        &  25.19 \\
			& 311d          &  11.76        &   3.36        &   7.15         &  9.72        &  11.12        &   8.17        &   8.98        &   9.34        &   9.30        &   9.32 \\
			& 328d          &  14.25        &  16.21        &  17.38         & 17.85        &  16.79        &  17.14        &  17.84        &  17.68        &  17.54        &  17.54 \\
			& 360d          &  54.41        &  54.72        &  56.63         & 54.59        &  55.38        &  54.56        &  55.44        &  55.80        &  55.61        &  55.57 \\
Barnase-barstar & 1b27          &    86.80       &   83.08      &    95.40       &   82.48       &   87.05       &   89.40      &    87.35       &   87.96       &   86.96       &   87.05 \\
				& 1b2s          &    67.80       &   66.33      &    68.81       &   71.75       &   70.03       &   71.47      &    72.32       &   71.93       &   72.25       &   72.12 \\
				& 1b2u          &    85.97       &   76.39      &    74.78       &   78.29       &   75.29       &   77.35      &    77.15       &   78.38       &   78.87       &   78.57 \\
				& 1b3s          &    48.41       &   56.58      &    49.61       &   46.02       &   49.87       &   48.17      &    53.44       &   49.38       &   49.07       &   49.25 \\
				& 1x1u          &   61.75     &     66.86      &    74.53       &   75.07     &     77.56        &  76.56      &    75.06     &     76.41      &    76.06     &     75.95\\
				& 1x1w          &    90.62       &   87.91      &    99.13       &   93.67       &   94.65       &   93.55      &    95.53       &   95.32       &   95.47       &   95.30 \\
				& 1x1x          &   115.79  &        110.40     &    110.45    &     112.19     &    113.51     &    114.45    &     115.30      &   114.65     &    114.62     &    114.65  \\
				& 1x1y          &    67.27       &   88.13      &    89.80       &   90.39       &   87.87       &   87.24      &    88.54       &   88.91       &   89.20       &   89.21 \\
				& 2za4          &    74.13       &   70.86      &    70.64       &   69.80       &   72.45       &   73.22      &    73.26       &   74.18       &   74.00       &   74.35 \\
RNA-peptide  & 1a1t\tnote{*}      &    			62.24     &     58.24    &      61.71     &     61.89    &      62.94    &      63.18   &       62.88     &     62.75     &     62.95     &     63.00 \\
& 1a4t\tnote{*}       &            72.37     &     72.44    &      69.76     &     69.94    &      71.19    &      72.46   &       72.41     &     72.39     &     72.24     &     72.27 \\
& 1biv\tnote{*}       &            41.80     &     40.73    &      42.07     &     44.90    &      45.66    &      44.70   &       44.69     &     44.75     &     44.86     &     44.76 \\
& 1exy\tnote{*}       &           178.36     &    178.17    &     178.16     &    176.29    &     178.70    &     176.91   &      177.52     &    177.50     &    177.70     &    177.36 \\
& 1g70                &           133.22     &    131.38    &     132.83     &    132.75    &     132.83    &     133.85   &      133.94     &    134.37     &    134.34     &    133.53 \\
& 1hji      &            46.78     &     46.06    &      49.80     &     50.70    &      51.51    &      51.26   &       51.21     &     51.42     &     51.17     &     51.23 \\
& 1i9f       &           -19.78     &    -22.55    &     -22.49     &    -19.44    &     -19.31    &     -19.20   &      -19.18     &    -19.18     &    -19.20     &    -19.22 \\
& 1mnb       &           126.65     &    129.00    &     127.74     &    126.95    &     127.80    &     127.82   &      128.30     &    128.15     &    128.15     &    128.20 \\
& 1nyb                &           -14.63     &    -13.07    &     -13.62     &    -12.58    &     -11.16    &     -13.04   &      -12.79     &    -12.45     &    -12.41     &    -12.61 \\
& 1qfq      &            18.09     &     19.84    &      16.64     &     20.52    &      21.19    &      20.03   &       22.60     &     20.66     &     20.19     &     20.32 \\
& 1ull\tnote{*}       &           -53.38     &    -58.11    &     -54.20     &    -53.61    &     -51.66    &     -53.28   &      -52.52     &    -52.76     &    -52.66     &    -52.75 \\
& 1zbn\tnote{*}       &           216.31     &    214.29    &     215.60     &    214.96    &     214.84    &     216.05   &      215.70     &    215.95     &    215.94     &    215.74 \\
& 2a9x      &           385.99     &    384.41    &     385.18     &    385.36    &     385.05    &     385.20   &      385.36     &    385.36     &    385.43     &    385.44 \\
& 484d\tnote{*}       &           129.72     &    129.65    &     133.42     &    132.09    &     131.34    &     132.03   &      132.79     &    133.20     &    133.23     &    133.38 \\       	      
         \bottomrule
\end{tabular}
	\begin{tablenotes}
	    \item [*] Results are significantly different (>50 kcal/mol) from those in  Ref. \cite{Harris:2013}.
	\end{tablenotes}
\end{threeparttable}
\end{table}

\begin{figure}[!tb]
	\centering
	\includegraphics[width=0.5\columnwidth]{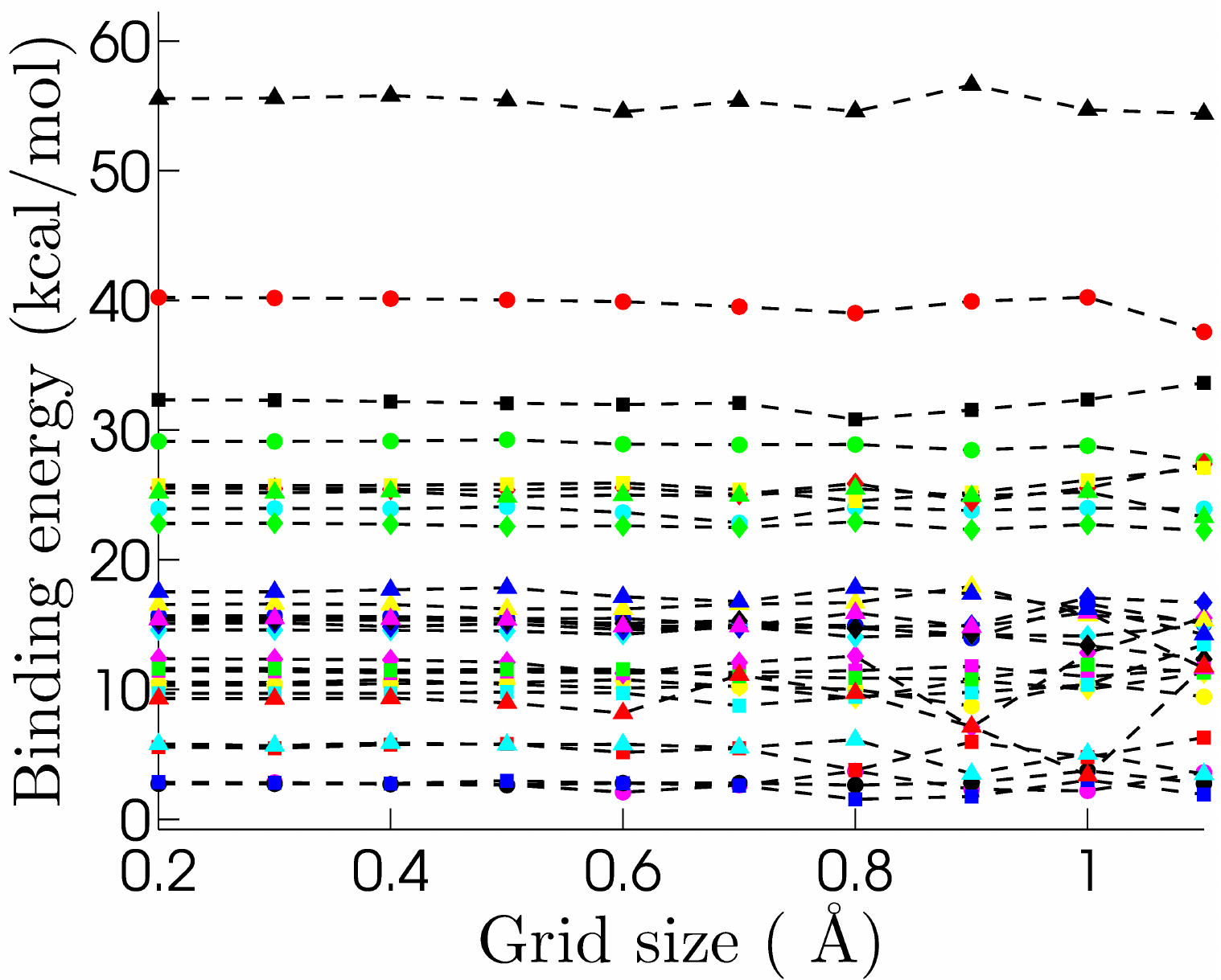}
	\caption{Binding electrostatic energy for  DNA-drug complexes with grid sizes from $\SI{0.2}{\angstrom}$ to $\SI{1.1}{\angstrom}$. The markers and PDBIDs are as follows yellow  circle  : 102d, magenta circle  : 109d, cyan    circle  : 121d, green   circle  : 127d, red     circle  : 129d, blue    circle  : 166d, black   circle  : 195d, yellow  diamond : 1d30, magenta diamond : 1d63, cyan    diamond : 1d64, green   diamond : 1d86, red     diamond : 1dne, blue    diamond : 1eel, black   diamond : 1fmq, yellow  square  : 1fms, magenta square  : 1jtl, cyan    square  : 1lex, green   square  : 1prp, red     square  : 227d, blue    square  : 261d, black   square  : 264d, yellow  triangle: 289d, magenta triangle: 298d, cyan    triangle: 2dbe, green   triangle: 302d, red     triangle: 311d, blue    triangle: 328d, black   triangle: 360d. }
	\label{fig.bd_d_d_line}
\end{figure}

%\begin{figure}[!tb]
%	\centering
%	\includegraphics[width=0.5\columnwidth]{binding_b-b.pdf}
%	\caption{Binding electrostatic energy for  barnase-barstar complexes with grid sizes from $\SI{0.2}{\angstrom}$ to $\SI{1.1}{\angstrom}$. The markers and PDBIDs are as follows yellow  circle : 1b27, magenta circle : 1b2s, cyan    circle : 1b2u, green   circle : 1b3s, red     circle : 1x1u, blue    circle : 1x1w, black   circle : 1x1x, yellow  diamond: 1x1y, magenta diamond: 2za4. }
%	\label{fig.bd_b_b_line}
%\end{figure}

%\begin{figure}[!tb]
%	\centering
%	\includegraphics[width=0.5\columnwidth]{binding_r-p.pdf}
%	\caption{Binding electrostatic energy for  RNA-peptide complexes with grid sizes from $\SI{0.2}{\angstrom}$ to $\SI{1.1}{\angstrom}$. The markers and PDBIDs are as follows yellow  circle  : 1a1t, magenta circle  : 1a4t, cyan    circle  : 1biv, green   circle  : 1exy, red     circle  : 1g70, blue    circle  : 1hji, black   circle  : 1i9f, yellow  diamond : 1mnb, magenta diamond : 1nyb, cyan    diamond : 1qfq, red     diamond : 1ull, green   diamond : 1zbn, blue    diamond : 2a9x, black   diamond : 484d. }
%	\label{fig.bd_r_p_line}
%\end{figure}

\begin{figure}[!tb]
	\begin{center}
		\includegraphics[width=0.50\columnwidth]{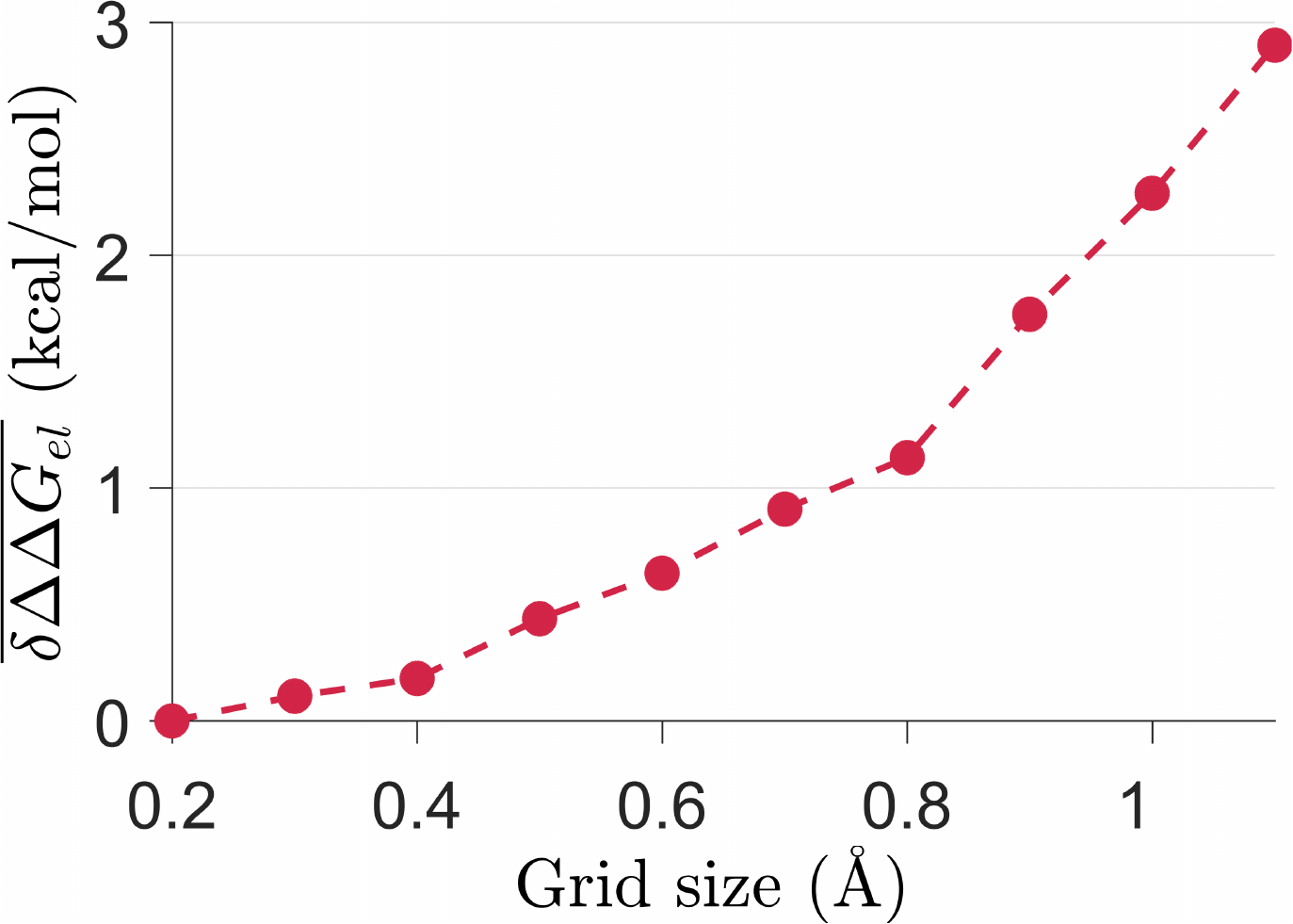}
		\caption{Averaged absolute error of the binding free energies for all the 51 complexes with mesh size refinements from $\SI{1.1}{\angstrom}$ to $\SI{0.2}{\angstrom}$.}
		\label{fig.err_bd}
	\end{center}
\end{figure}

Similar to the study of the convergence of $\Delta G_{\text{el}}$, we correlate the binding free energy calculated at the finest grid spacing, $h=\SI{0.2}{\angstrom}$, and ones estimated at coarser mesh sizes, $h=\SI{0.3}{\angstrom},\cdots,\SI{1.1}{\angstrom}$. Figure \ref{fig.bd} illustrates these relationships with the regression lines whose parameters are revealed in Table \ref{tab.r2_bd}. Since the previous discussion confirms MIBPB solver can produce very good R-squared values even at very coarse grid spacings, it is interesting to explore whether a similar behavior can be found for binding energy estimation. Indeed, the PB  binding energy estimation behaves the same as the PB solvation calculation in our MIBPB technique. Specifically, $R^2$ is always $1$ at the fine mesh, $h=\SI{0.3}{\angstrom}$. Moreover, these values are still satisfactory at relatively coarser mesh sizes. For example, at the grid spacing of $h=1.1$, the $R^2$ and slope of the regression line for DNA-drug, barnase-barstar, and RNA-peptide complexes are, respectively, (0.9747,1.0081), (0.8002,0.8187),  and (0.9998, 0.9937). In contrast, the R-squared values reported in Ref. \cite{Harris:2013}, computed between $\SI{0.3}{\angstrom}$ and $\SI{1.0}{\angstrom}$, are unacceptable for SESs, and usually less than $0.62$. 
Our statistical measures strongly support the reliable binding energy prediction of our solver at coarse grid sizes. Table \ref{tab.binding_eng} displays the binding free energy for all complexes with different grid spacings. As can be seen from Table \ref{tab.binding_eng}, the difference between binding energies at coarse meshes and the finest mesh, $h=\SI{0.2}{\angstrom}$, is mostly less than 10 kcal/mol for all complexes.

% \ref{fig.bd_b_b_line}, and \ref{fig.bd_r_p_line}
The trend of binding free energy at different grid spacings can be seen clearly in  Figs. \ref{fig.bd_d_d_line} which plots $\Delta\Delta G_{\text{el}}$ against grid sizes varying between $\SI{0.2}{\angstrom}$ and $\SI{1.1}{\angstrom}$ for DNA-drug complexes. Similar figures for barnase-barstar and RNA-peptide complexes can be referred to Figs. S1 and S2 in the Supporting Information. Based on these figures, our solver can rank the binding free energy for DNA-drug complexes at grid spacing of $\SI{0.6}{\angstrom}$, barnase-barstar complexes at grid spacing of $\SI{0.6}{\angstrom}$, and RNA-peptide complexes at significantly coarse grid spacing of $\SI{1.1}{\angstrom}$. To further assess the reliable estimates of binding energy of our MIBPB solver, we consider the absolute difference between results computed at a coarser grid spacing  and the finest grid spacing defined by
\begin{align}\label{error.bd}
\delta\Delta\Delta G_{\text{el}}=\left|\Delta\Delta G_{\text{el},h} - \Delta\Delta G_{\text{el},h=0.2}\right|.
\end{align}

Figure \ref{fig.err_bd} plots the averaged absolute errors, $\overline{\delta\Delta\Delta G_{\text{el}}}$, i.e., the average of absolute errors defined in Eq. \eqref{error.bd} over all 51 complexes, at different mesh sizes. It is seen  that even the use of grid spacing of $\SI{0.7}{\angstrom}$ still delivers an averaged binding calculation error under 1 kcal/mol for this set of complexes.
Therefore, we can draw a conclusion that the common use of grid size being $\SI{0.5}{\angstrom}$ is still adequate for predicting the binding energy free without producing a misleading result.

\begin{figure}[!tb]
		\begin{center}
			\includegraphics[width=0.60\columnwidth]{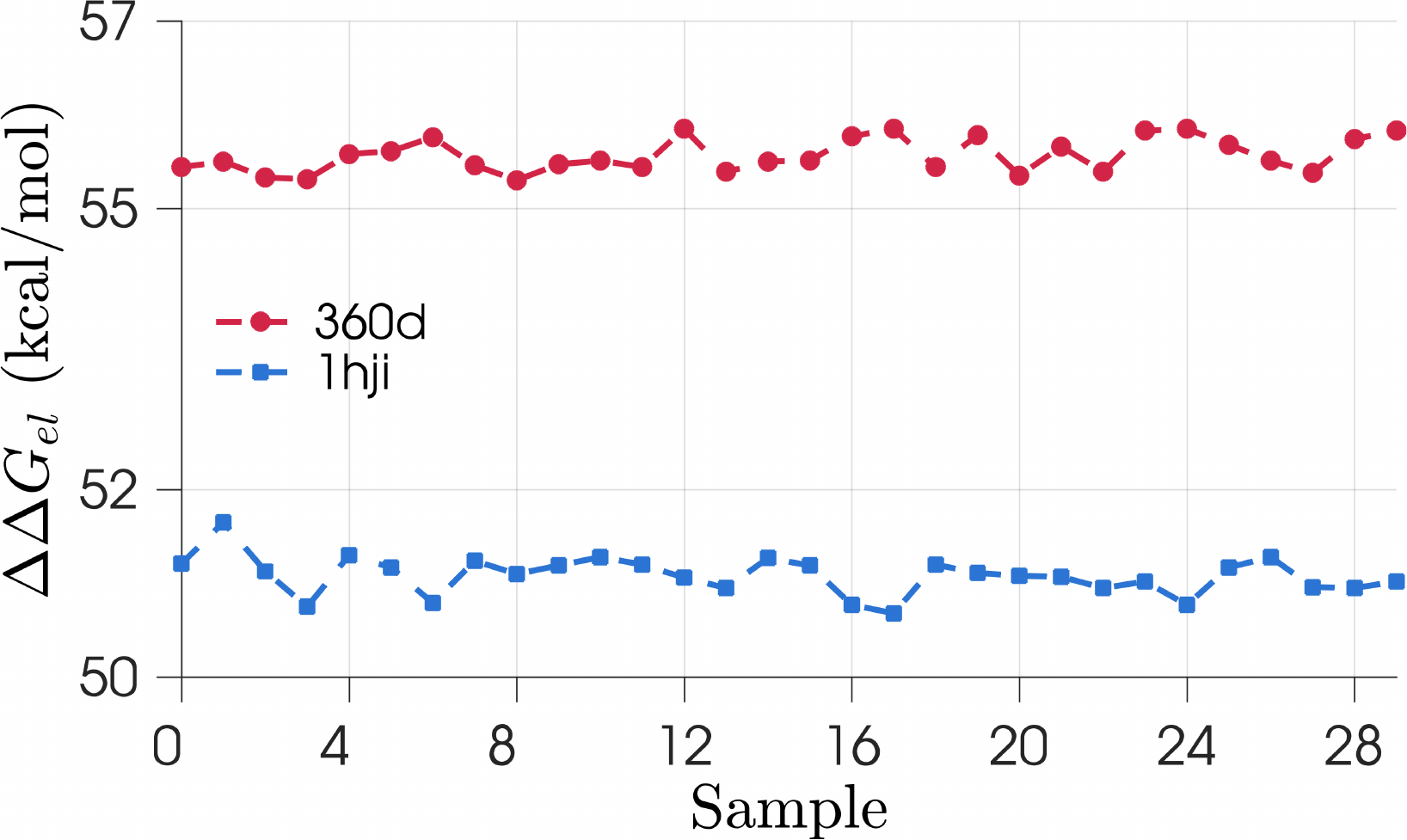}
			\caption{Binding energies of two complexes PDB IDs: 360d (marked by circle) and 1hji (marked by square) at 30 different grid positions.}
			\label{fig.std_dev_bd}
		\end{center}
	\end{figure}

	 Grid positioning error is another feature to validate the robustness and accuracy of a PB solver. To examine such numerical error for our MIBPB solver, we consider two protein complexes with PDBIDs: 360d and 1hji.  To estimate the standard deviation, $\sigma_{\rm bd}$ in $\Delta\Delta G_{\rm el}$, we randomly generate 29 grid positions around the initial origin with the amplitude of the random seed being $\pm 0.5h$, where $h=0.5~\text{\AA~}$ is the grid spacing.  Then $\Delta\Delta G_{\rm el}$ is evaluated at all of the 30 grid positions.   Figure \ref{fig.std_dev_bd} plots electrostatic binding energies at 30 distinct samples of grid positions, including the original one marked by Sample 0 on the graph. The $\sigma_{\rm bd}$ values of complexes  360d and 1hji are found to be 0.18 and 0.21, respectively. These results indicate that the MIBPB solver is not sensitive to grid position.

Note that in Table \ref{tab.binding_eng}, some binding energies obtained in our calculations, all from RNA-peptide complexes, differ significantly  from those reported in Ref. \cite{Harris:2013}, i.e., more than 50 kcal/mol at the finest grid spacing, under the same parametrization and   data inputs. However,   overall, our results have a good agreement with those in  Ref. \cite{Harris:2013} with $R^2=0.9135$ for RNA-peptide complexes.   To further support our calculations, we have employed PBSA, Delphi, and APBS for electrostatic energy calculations at the grid size of 0.2 \AA .  We note that results obtained from these solvers are in excellent agreements, i.e., $R^2>0.98$, with ours.
The electrostatic energies calculated by PBSA, Delphi, and APBS solvers are listed in Table S2 of Supporting Information.

\section{Concluding remarks}

 Poisson-Boltzmann (PB) theory is an established model for biomolecular electrostatic analysis and has been widely used in electrostatic solvation   $\Delta G_{\text{el}}$ and binding energy $\Delta\Delta G_{\text{el}}$  estimations.  However,  doubt   has been cast  on the validity of  the commonly used grid spacing of 0.5 \AA~  for producing converged estimates of $\Delta\Delta G_{\text{el}}$ due to the unacceptable errors observed in the  calculation using the solvent excluded surface (SES) and the  adaptive Cartesian grid (ACG) finite difference PB equation solver \cite{Harris:2013}.  Three sets of biomolecular complexes, namely,   DNA-drug complexes, barnase-barstar complexes, and RNA-peptide complexes,  are employed in the  study. The discrepancies between results obtained from different surface definitions were also utilized to support the  general pessimism for the PB methodology.

In this work, we employ the MIBPB software \cite{DuanChen:2011a,Geng:2007a} to estimate electrostatic solvation free energy, $\Delta G_{\text{el}}$, and binding free electrostatic energy, $\Delta\Delta G_{\text{el}}$, for the three sets of biomolecular complexes used  in Ref. \cite{Harris:2013}. The popular SES is adopted in the present work. In our $\Delta G_{\text{el}}$ estimation, the averaged relative absolute error computed at a relatively coarse grid size of $\SI{1.1}{\angstrom}$ against the finest grid size of 0.2 \AA~ over 153 studied biomolecules is less than 0.31\%. The same error obtained at the grid size of $\SI{1.0}{\angstrom}$ is less than 0.2\%. These results indicate the reliability of using the MIBPB solver at the grid spacing of $\SI{1.0}{\angstrom}$ or even $\SI{1.1}{\angstrom}$ for electrostatic solvation analysis.  The robustness and accuracy of MIBPB solver for estimates of $\Delta G_{\text{el}}$ have been reported for 24 proteins in the literature \cite{DuanChen:2011a,Geng:2007a}. This characteristics has been confirmed again in the present work for  DNA-drug complexes, barnase-barstar complexes, and RNA-peptide complexes.

The well-converged $\Delta G_{\text{el}}$ produced by our solver enables a promising performance in predicting $\Delta\Delta G_{\text{el}}$ at a coarse grid spacing. Indeed, numerical estimates of $\Delta\Delta G_{\text{el}}$ in the current work reveals that $\Delta\Delta G_{\text{el}}$ obtained at a $\SI{1.1}{\angstrom}$ grid spacing mostly differ by less than 10 kcal/mol from that achieved by using a $\SI{0.2}{\angstrom}$ grid spacing. Moreover, MIBPB solver conducted at grid size of $\SI{0.6}{\angstrom}$ perfectly produces a well-converged $\Delta\Delta G_{\text{el}}$, and qualitatively ranks the complexes in term of  their binding free energies. Therefore, the current results support an opinion that the widely used grid size of $\SI{0.5}{\angstrom}$ can give  reliable and accurate enough predictions of both electrostatic free energy \cite{nicholls2008predicting,shivakumar2010prediction} and binding free energy.

  To develop highly accurate, robust and reliable PB solvers for biomolecular electrostatics, it is crucial to validate one's numerical methods by  appropriate norms and against realistic problems. We emphasize that as an elliptic interface problem, it is important to measure the convergence of PB solvers in the $L_{\infty}$ norm, or maximum absolute error, because integral norms, such as $L_1$ and $L_2$, are insensitive to the performance of numerical methods near the interface. Additionally, the convergence should be tested by  solving the PB equation, rather than by calculating the solvation free energy. Finally,  validation should be carried out by using the SESs of proteins, rather than smooth surfaces, such as a sphere.     

\begin{suppinfo}
%	\begin{itemize}
%		\item 
		Electrostatic free energies calculated by different solvers, namely MIBPB, DELPHI, PBSA and APBS; Coulombic binding energies; binding energy plots for barnase-barstar and RNA-peptide complexes (filename: \url{jctc_si_bdenergy.pdf}).
		%\item  Python code used to calculate Coulombic binding energies (filename:\url{coulomb.py}).		
%	\end{itemize}

\end{suppinfo}		

\begin{acknowledgement}
This work was supported in part by NSF Grant Nos. IIS- 1302285 and
DMS-1160352,   NIH Grant No. R01GM-090208 and  MSU Center for Mathematical Molecular Biosciences Initiative. %And we thank the ICER high performance computing center for the computing resources. 
DDN and GWW thank the Mathematical Biosciences Institute  for its hospitality and support during their visit in Ohio State University, where this manuscript was finalized.
This manuscript was reviewed by Professors Emil Alexov and Ray Luo prior its submission.  

\end{acknowledgement}

%\begin{suppinfo}
%
%This will usually read something like: ``Experimental procedures and
%characterization data for all new compounds. The class will
%automatically add a sentence pointing to the information on-line:
%
%\end{suppinfo}
\bibliographystyle{abbrv}
\bibliography{refs}

\newpage

\begin{figure}[!tb]
	\centering
	\includegraphics[width=1.0\columnwidth]{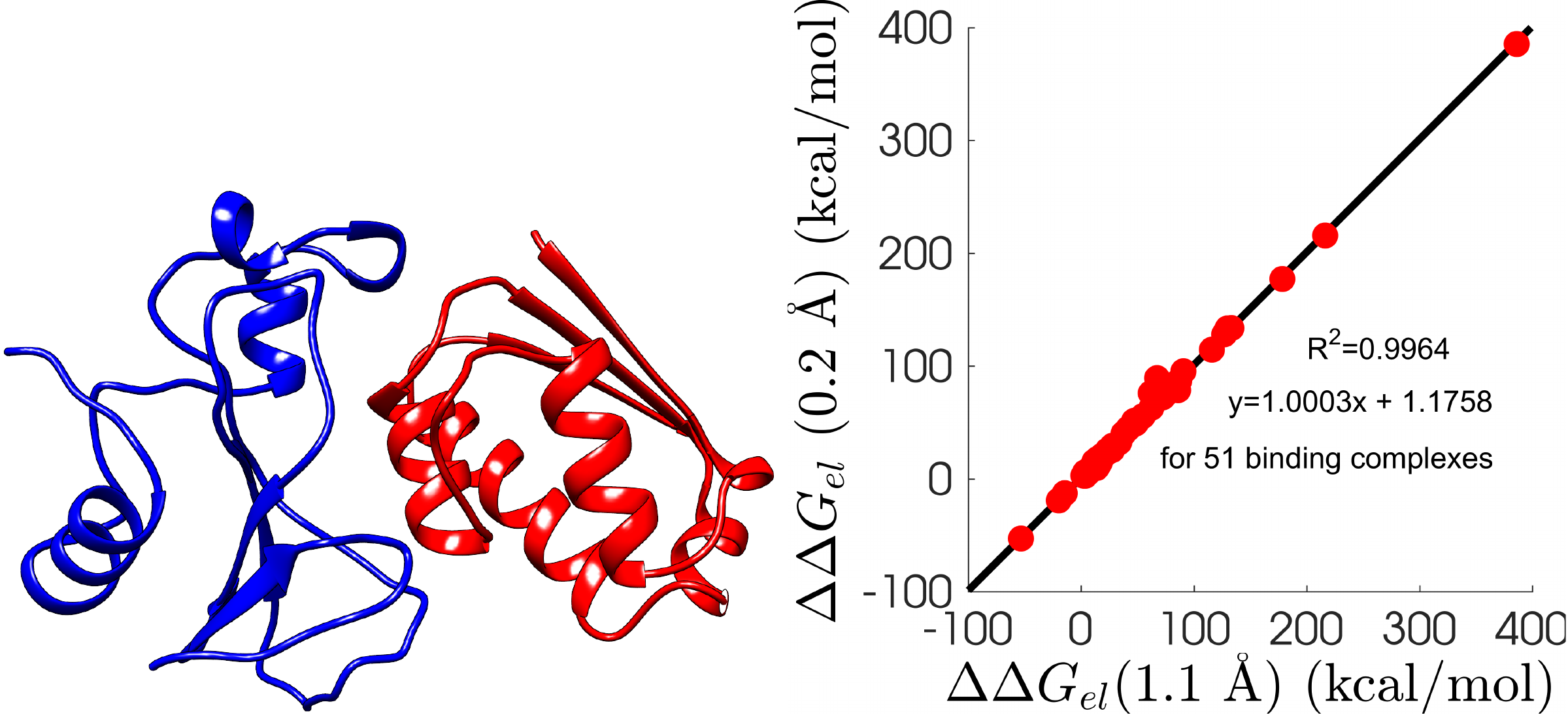}
\end{figure}
\end{document}